\documentclass[]{wiley-article} 
\usepackage[authoryear]{natbib}
\makeatletter
\renewcommand{\@biblabel}[1]{#1.}
\makeatother

\usepackage{physics}
\usepackage{placeins}
\usepackage{siunitx}
\usepackage[noabbrev,capitalise]{cleveref}
\newcommand{\cxv}{\vb{X}}
\newcommand{\ct}{\mathrm{T}}
\newcommand{\cx}{\mathrm{X}}
\newcommand{\cy}{\mathrm{Y}}
\newcommand{\cz}{\mathrm{Z}}
\newcommand{\mean}[1]{\left\langle{#1}\right\rangle}

\newcommand{\matder}{\mathrm{D}_t}

\papertype{Research Article}

\title{Impact of Small-Scale Gravity Waves on Tracer Transport}

\author[1]{Irmgard Knop}
\author[1]{Stamen Dolaptchiev}
\author[1]{Ulrich Achatz}

\affil[1]{Institute for Atmosphere and Environment, Goethe University Frankfurt, Germany}

\corraddress{Irmgard Knop, Institute for Atmosphere and Environment, Goethe University Frankfurt, Altenhöferallee 1, D-60438 Frankfurt am Main, Germany}
\corremail{knop@iau.uni-frankfurt.de}

\fundinginfo{Deutsche Forschungsgemeinschaft, Grant/Award Number: 428312742, 274762653}

\runningauthor{Knop et al.}

\begin{document}

\begin{frontmatter}
\maketitle

\begin{abstract}
Small-scale gravity waves, with horizontal wavelengths of up to a few hundred kilometers and vertical wavelengths of up to a few kilometers, play a crucial role in atmospheric tracer transport. However, their effects remain unresolved in climate models and must be parameterized. This study investigates how gravity waves influence large-scale tracer distributions, utilizing a multiple-scale analysis to systematically identify the governing terms of gravity wave-induced tracer fluxes. The analysis reveals both leading-order and next-order impacts: the former being the inertia-gravity wave-induced tracer Stokes drift, which acts perpendicular to both the large-scale tracer gradient and the wave number, while the latter becomes significant at lower latitudes where Coriolis effects diminish. A numerical framework is developed to incorporate these fluxes into a gravity wave parameterization model, potentially enhancing climate model accuracy without requiring explicit resolution of small-scale wave dynamics. Model validation against high-resolution wave-resolving simulations confirms the effectiveness of this approach. By improving the representation of gravity wave-induced tracer transport, this research advances the accuracy of climate simulations, particularly in their depiction of microphysics and radiative processes.

\end{abstract}
\end{frontmatter}

\section{Introduction}

Accurately simulating the three-dimensional distribution of tracers is crucial for climate modeling, as these tracers have significant radiative effects on Earth's climate. In the zonal mean, large-scale tracer transport—such as that of ozone and water vapor—is primarily governed by thermally and wave-driven global circulations \citep{Butchart2014Brewer,Brewer1949,Dobson1956}. Small-scale processes, such as gravity waves and turbulence, indirectly influence tracer distributions by driving and modifying these circulations. However, they can also directly affect tracer transport and mixing. These gravity waves, with horizontal and vertical wavelengths of up to a few hundred and a few kilometers, respectively, remain unresolved in climate models and are small compared to synoptic-scale processes, which extend over several thousand kilometers. Since climate models do not resolve these small-scale processes, their effects must be represented through parameterization. 

Previous studies have shown that gravity waves can significantly impact tracer transport \citep{Xu2002StudyAndApplication,Walterscheid1989GravityWaveFluxes,Liu2004VerticalDynamical}. An analysis by \citet{heller2016mountain} shows that mountain waves can generate water vapor fluxes large enough to alter radiative forcing by more than \SI{1}{\watt\per\meter}. Although weaker than turbulent fluxes \citep[see e.g.][]{wu2016observations}, these fluxes are substantial enough to warrant parameterization. A parameterization of these effects has been proposed by \citet{Gardner2019Parameterizing}, focusing on vertical mixing arising from wave breaking and propagating waves. While comparison with observations proves difficult, their parameterization has been shown to be a more physical approach when accounting for gravity wave-induced tracer transport. However, a systematic analysis of the governing equations is still lacking, leaving the full nature of gravity wave-induced tracer fluxes uncharacterized. A more refined parameterization is needed to capture these effects comprehensively.

This study aims to address this gap by systematically deriving the key terms that govern how small-scale gravity waves influence large-scale tracer transport. Furthermore, it develops a framework for incorporating these effects into gravity wave parameterization models. To achieve this, we employ a multiple-scale analysis to separate and analyze the interactions between large-scale tracer transport and small-scale gravity waves, identifying the dominant contributions.

Improving the representation of gravity-wave induced tracer transport can enhance climate model accuracy. The framework developed in this study allows for high-accuracy tracer modeling in the presence of gravity waves without explicitly resolving them, significantly improving computational efficiency. Additionally, this research provides a foundation for quantifying the role of non-breaking gravity waves in tracer transport.

This paper is structured as follows. \Cref{sec:theory} derives the inertial and mid-frequency gravity wave-induced tracer transport using a multiple-scale analysis of the governing equations. Using a WKB ansatz, we obtain expressions for wave-induced tracer fluxes in terms of wave amplitudes. We then show how these fluxes can be computed from the phase-space wave action density, a quantity used in gravity wave ray tracing models. The most important equations are summarized in \cref{sec:summary-equations}. \Cref{sec:numerical-model} describes the extension of the parameterization model MS-GWaM \citep[see e.g.][]{Boloni2016Interaction, Wei2019Efficient,Jochum2025} to include these fluxes and its coupling to an idealized flow solver. We validate the model by comparing its results to wave-resolving reference simulations. Finally, we summarize and conclude our results in \cref{sec:summary}.

\section{Theory of Gravity Wave-Induced Tracer Transport} \label{sec:theory}

This section provides the theoretical basis for understanding the impact of non-breaking, small-scale gravity waves on large-scale tracer transport through gravity wave tracer fluxes. 
We begin by presenting the governing equations and characteristic scales of gravity waves, then derive expressions for the flux convergences they induce. 
Using the WKB ansatz, we reformulate these flux convergences in terms of the phase-space wave action density, which serves as the basis for their parameterization.

\subsection{Multiple-scale analysis of the governing equations}

In this section, we first present the governing equations, including the tracer advection equations. 
Using the characteristic scales of small-scale gravity waves, we non-dimensionalize these equations and separate terms by order of magnitude in a scale-separation parameter. This approach allows us to identify the terms that describe the impact of gravity waves on large-scale tracer transport. 

The advection of a unit-less passive tracer mixing ratio \(\psi\), neglecting diffusion, is given by 
\begin{equation}
    \label{eq:tracer-eq-dim-form}
    \matder\psi = 0\;.
\end{equation}
The material derivative is defined as \(\matder=\partial_t+\vb{v}\cdot\nabla\), 
where \(\vb{v}=(u,v,w)^{\mathrm{T}}\) with \(u\), \(v\), and \(w\) representing the wind components in the \(x\)-, \(y\)-, and \(z\)-directions, respectively. The gradient operator is defined as \(\nabla = \partial_x\vb{e}_x + \partial_y\vb{e}_y + \partial_z\vb{e}_z\) with unit vectors \(\vb{e}_x\), \(\vb{e}_y\), and \(\vb{e}_z\).
The wind velocity \(\vb{v}\), potential temperature \(\theta\), and Exner pressure \(\pi\) are governed by the equations of motion for a dry, inviscid atmosphere \citep{Achatz2022Chapter10}
\begin{align}
    \label{eq:hor-mom-eq-dim-form}
    \matder\vb{u} + f\vb{e}_z\times\vb{u} &= -c_p\theta\nabla_h\pi \;, \\
    \label{eq:vert-mom-eq-dim-form}
    \matder w &= -c_p\theta\partial_z\pi - g\;, \\
    \label{eq:pot-temp-eq-dim-form}
    \matder\theta &= 0 \;, \\
    \label{eq:exner-eq-dim-form}
    \matder\pi + \frac{R}{c_V}\pi\nabla\cdot\vb{v} &= 0\;.
\end{align}
Here, \(\vb{u}\) represents the horizontal wind components, \(c_p\) and \(c_V=c_p-R\) are the specific heat capacities at constant pressure and volume, respectively, \(R\) is the ideal gas constant of dry air, \(g\) is the gravitational acceleration, and \(f\) is the constant Coriolis frequency. Furthermore,  we define the horizontal gradient operator \(\nabla_h=\partial_x\vb{e}_x + \partial_y\vb{e}_y\).

To facilitate the multiple-scale analysis of \cref{eq:tracer-eq-dim-form,eq:hor-mom-eq-dim-form,eq:vert-mom-eq-dim-form,eq:pot-temp-eq-dim-form,eq:exner-eq-dim-form}, we first non-dimensionalize these equations.
Following an approach equivalent to \citet{Achatz2017Interaction}, we define the necessary scales by assuming a hydrostatic reference atmosphere with potential temperature, density, and Exner pressure profiles \(\overline{\theta}(z)\), \(\overline{\rho}(z)\), and \(\overline{\pi}(z)\), respectively.
The Brunt-Väisälä frequency is given by \(N=\sqrt{(g/\overline{\theta})\dd_{z}\overline{\theta}}\). 
Depending on the atmospheric stratification, the ratio between the Coriolis frequency and the Brunt-Väisälä frequency is given by \(f/N=\mathcal{O}\left[\varepsilon^{(5-\delta)/2}\right]\), where \(\delta=0\) or \(\delta=1\) for a moderately strong or weakly stratified atmosphere, respectively. For \(\delta=1\), the weak stratification represents conditions in the troposphere and mesosphere, whereas \(\delta=0\) corresponds to the strong stability of the stratosphere. Here, \(\varepsilon=\widetilde{L}_w/\widetilde{L}=\widetilde{H}_w/\widetilde{H}=\widetilde{T}_w/\widetilde{T}\) is a scale separation parameter taken to be identical to the Rossby number \citep{Achatz2017Interaction},
\begin{equation}
    \varepsilon = \frac{\widetilde{U}}{f\widetilde{L}} = \frac{1}{f\widetilde{T}} = \mathcal{O}\left(10^{-1}\right) \ll 1 \;,
\end{equation}
where \(\widetilde{U}\), \(\widetilde{L}\), \(\widetilde{H}\), and \(\widetilde{T}\) denote characteristic synoptic-scale horizontal velocity, length, height, and time scales, respectively (see \cref{tab:scaling-parameters}), while \(\widetilde{L}_w\), \(\widetilde{H}_w\), and \(\widetilde{T}_w\) represent the characteristic length, height, and times scales of gravity waves. These synoptic scales provide the basis for non-dimensionalizing \cref{eq:tracer-eq-dim-form,eq:hor-mom-eq-dim-form,eq:vert-mom-eq-dim-form,eq:pot-temp-eq-dim-form,eq:exner-eq-dim-form}, yielding the system
\begin{align}
    \label{eq:tracer-eq-nondim-form}
    \matder\psi &= 0\;, \\
    \label{eq:hor-mom-eq-nondim-form}
    \varepsilon^{2+\delta}\left(\matder\vb{u} + f_0\vb{e}_z\times\vb{u}\right) &= -c_p\theta\nabla_h\pi \;, \\
    \label{eq:vert-mom-eq-nondim-form}
    \varepsilon^7\matder w &= -c_p\theta\partial_z\pi - 1\;, \\
    \label{eq:pot-temp-eq-nondim-form}
    \matder\theta &= 0 \;, \\
    \label{eq:exner-eq-nondim-form}
    \matder\pi + \frac{R}{c_V}\pi\nabla\cdot\vb{v} &= 0\;,
\end{align}
where \(f_0=1\) serves as a non-dimensional placeholder for the Coriolis parameter. Here, \(\matder = \partial_t + \vb{v}\cdot\nabla\) denotes the non-dimensional material derivative.

\begin{table}[bt]
\caption{Summary of all scaling parameters, with \(\delta=0\) and \(1\) for a strongly and weakly stratified atmosphere, respectively, a scaling parameter \(\varepsilon=\mathcal{O}(1/10)\), and a typical atmosphere temperature \(T_{00}\). Parameters in accordance to \citet{Achatz2023MultiScaleDynamics}.}\label{tab:scaling-parameters}
\centering
\begin{threeparttable}
\begin{tabular}{c c c}
\headrow
\thead{Name} & \thead{Symbol} & \thead{Scale size} \\
Horizontal scale & \(\widetilde{L}\) & \(\varepsilon^{(2+\delta)/2}\sqrt{RT_{00}}/f\)  \\
Vertical scale & \(\widetilde{H}\) & \(\varepsilon^{7/2}\sqrt{RT_{00}}/f\)  \\
Time scale & \(\widetilde{T}\) & \(1/f\)  \\
Horizontal velocity & \(\widetilde{U}\) & \(\widetilde{L}/\widetilde{T}=\varepsilon^{(2+\delta)/2}\sqrt{RT_{00}}\)  \\
Vertical velocity & \(\widetilde{W}\) & \(\widetilde{H}/\widetilde{T} =\varepsilon^{7/2}\sqrt{RT_{00}}\)  \\
Characteristic tracer mixing ratio & \(\widetilde{\Psi}\) & \(1\) \\
\hline
\end{tabular}
\end{threeparttable}
\end{table}

As solutions to \cref{eq:tracer-eq-nondim-form,eq:hor-mom-eq-nondim-form,eq:vert-mom-eq-nondim-form,eq:pot-temp-eq-nondim-form,eq:exner-eq-nondim-form}, we consider a superposition of expansions in terms of the scale-separation parameter for the reference atmosphere, the large-scale flow, and gravity waves described by the WKB ansatz. For clarity, we restrict the main presentation to a locally monochromatic, large-amplitude wave in a weakly stratified atmosphere (\(\delta=1\)). Here, large-amplitude means that the wave-induced wind field is of the same order of magnitude as the large-sacle wind field. A more general expansion, including a spectrum of gravity waves with higher harmonics under both strong and weak stratification, is provided in \cref{sec:appendix-full-expansion}. The variables thus take the form
\begin{align}
    \label{eq:tracer-decomp}
    \psi(\vb{x},t) =& \sum_{m=0}^{\infty}\varepsilon^m\mean{\Psi}^{(m)}(\cxv,\ct) 
    + \mathfrak{R}\sum_{m=0}^{\infty}\varepsilon^m\hat{\Psi}^{(m)}(\cxv,\ct)e^{i\phi(\cxv,\ct)/\varepsilon} \;, \\
    \label{eq:wind-decomp}
    \vb{v}(\vb{x},t) =& \sum_{m=0}^{\infty} \varepsilon^m\mean{\vb{V}}^{(m)}(\cxv,\ct) + \mathfrak{R}\sum_{m=0}^{\infty}\varepsilon^{m}\hat{\vb{V}}^{(m)}(\cxv,\ct)e^{i\phi(\cxv,\ct)/\varepsilon} \;, \\
    \label{eq:pot-temp-decomp}
    \begin{split}
        \theta(\vb{x},t) =& \sum_{m=0}^{1}\varepsilon^m\overline{\Theta}^{(m)}(\cz) + \varepsilon^{2}\sum_{m=0}^{\infty}\varepsilon^{m}\mean{\Theta}^{(m)}(\cxv,\ct)  
        + \varepsilon^{2}\mathfrak{R}\sum_{m=0}^{\infty}
        \varepsilon^{m}\hat{\Theta}^{(m)}(\cxv,\ct)e^{i\phi(\cxv,\ct)/\varepsilon} \;, 
    \end{split}  \\
    \label{eq:exner-decomp}
    \begin{split}
        \pi(\vb{x},t) =& \sum_{m=0}^{1}\varepsilon^m\overline{\Pi}^{(m)}(\cz) + 
        \varepsilon^{2}\sum_{m=0}^{\infty}\varepsilon^m\mean{\Pi}^{(m)}(\cxv,\ct)  
        + \varepsilon^{3}\mathfrak{R}\sum_{m=0}^{\infty}\varepsilon^m\hat{\Pi}^{(m)}(\cxv,\ct)e^{i\phi(\cxv,\ct)/\varepsilon} \;,
    \end{split}
\end{align}
The Exner pressure and potential temperature profiles of the reference atmosphere are denoted by \(\overline{\Pi}^{(m)}\) and \(\overline{\Theta}^{(m)}\), respectively, while large-scale variables are represented using angle brackets. The remaining variables, proportional to \(e^{i\phi/\varepsilon}\), correspond to wave contributions.
To emphasize the slow variation of the large-scale flow, wave amplitudes, and local wavenumbers \(\vb{k}=\nabla\phi/\varepsilon=\nabla_{\cxv}\phi\) and frequencies \(\omega=-\partial_t\phi/\varepsilon=-\partial_{\ct}\phi\), we introduce the slow variables \((\cxv,\ct)=\varepsilon(\vb{x},t)\). 

Before proceeding with the multiple-scale analysis of the tracer advection equation, we briefly outline how we handle the \(\varepsilon\)-expansion of the large-scale and background atmosphere variables and introduce a compact notation, which will be used throughout \cref{sec:theory}. 
\Cref{eq:tracer-decomp,eq:wind-decomp,eq:pot-temp-decomp,eq:exner-decomp} present the full expansion of the large-scale and reference atmosphere variables in \(\varepsilon\). In a rigorous asymptotic expansion of \cref{eq:tracer-eq-nondim-form,eq:hor-mom-eq-nondim-form,eq:vert-mom-eq-nondim-form,eq:pot-temp-eq-nondim-form,eq:exner-eq-nondim-form}, we would account for the different orders of magnitude in \(\varepsilon\) for these variables, as done in \cite{Achatz2023MultiScaleDynamics}. However, since the equations in this paper are designed for numerical implementation, we retain the full expansion rather than isolating specific orders in \(\varepsilon\). This approach is necessary, because, in numerical models, the large-scale variables represent the full resolved flow without any expansion in \(\varepsilon\). To streamline notation, we introduce a compact representation for the large-scale variables, exemplified here for the large-scale tracer
\begin{align}
    \sum_{m=0}^{\infty}\varepsilon^m\mean{\Psi}^{(m)} \equiv \mean{\underline{\Psi}} \;,
\end{align}
with equivalent notation for the large-scale winds, Exner pressure, and potential temperature. Additionally, the Exner pressure and potential temperature of the reference atmosphere are denoted by 
\begin{align}
    \sum_{m=0}^{1}\varepsilon^m\overline{\Pi}^{(m)} \equiv \overline{\underline{\Pi}} 
    \quad \text{and} \quad 
    \sum_{m=0}^{1}\varepsilon^m\overline{\Theta}^{(m)} \equiv \overline{\underline{\Theta}} \;,
\end{align}
respectively.

We now proceed with the multiple-scale analysis of the tracer advection equation, aiming to identify the dominant terms that describe the impact of gravity waves on large-scale tracer transport. 
By substituting the decompositions from \cref{eq:tracer-decomp,eq:wind-decomp} into \cref{eq:tracer-eq-nondim-form} and sorting terms by powers of \(\varepsilon\) and the phase factor \(e^{i\phi/\varepsilon}\), we find that the first nonzero tracer wave amplitude, \(\hat{\Psi}^{(1)}\), is related to the leading-order wind wave amplitude by 
\begin{equation}
    \label{eq:amplitude-eq-full-gamma0}
    i\hat{\omega}\hat{\Psi}^{(1)} = 
    \hat{\vb{V}}^{(0)}\cdot\nabla_{\cxv}\mean{\underline{\Psi}}\;,
\end{equation}
where \(\hat{\omega}=\omega-\vb{k}_{h}\cdot\mean{\underline{\vb{V}}}\) is the intrinsic frequency. This result shows that the gradients in the large-scale tracer field induce oscillations in the leading-order tracer perturbation field that share the same frequency as the wind fluctuations. 
The next-to-leading-order tracer amplitudes are given by the relation 
\begin{equation}
    \label{eq:amplitude-eq-full-gamma1}
    i\hat{\omega}\hat{\Psi}^{(2)} = \left(\partial_{\ct} + 
    \mean{\underline{\vb{U}}}\cdot\nabla_{\cxv,h}\right)
    \hat{\Psi}^{(1)} + \hat{\vb{V}}^{(1)}\cdot\nabla_{\cxv}\mean{\underline{\Psi}}\;.
\end{equation}
These results also account for the fact that the leading-order large-scale vertical wind, \(\mean{W}^{(0)}=0\), vanishes, as demonstrated in \citet{Achatz2023MultiScaleDynamics}.
Using the large-scale terms and the relations above, we obtain the following equation for the large-scale tracer mixing ratio
\begin{equation}
    \label{eq:large-scale-eq-full}
    \begin{split}
        \varepsilon
        \left(\partial_{\ct} 
            + 
            \mean{\underline{\vb{V}}}\cdot\nabla_{\cxv}
        \right)
            \mean{\underline{\Psi}} = - \frac{1}{2}\sum_{m=0}^{\infty}\sum_{n=1}^{\infty}&\varepsilon^{(m+n+1)}\mathfrak{R}\left(\hat{\vb{V}}^{(m)}\cdot\nabla_{\cxv}\hat{\Psi}^{(n)^*}\right)
        +\varepsilon^{(m+n)}\mathfrak{I}\left(\hat{\vb{V}}^{(m)}\cdot\vb{k}\hat{\Psi}^{(n)^*}\right) \;.
    \end{split}
\end{equation}
Additionally, using the wind amplitude relations derived from the Exner pressure equation (\cref{eq:exner-eq-nondim-form})
\begin{align} \label{eq:wind0}
    i\vb{k}\cdot\hat{\vb{V}}^{(0)} 
    &= 0\;, \\ \label{eq:wind1}
    i\vb{k}\cdot\hat{\vb{V}}^{(1)} 
    &= -\frac{1}{
        \overline{P}
        }
    \nabla_{\cxv}\cdot\left(
        \overline{P}
        \hat{\vb{V}}^{(0)}\right) \;, \\ 
    i\vb{k}\cdot\hat{\vb{V}}^{(2)} 
    &= -\frac{1}{
        \overline{P}
        }
    \nabla_{\cxv}\cdot\left(
        \overline{P}
        \hat{\vb{V}}^{(1)}\right)  \;, \label{eq:solenoid}
\end{align}
where \(\overline{P}=\overline{\rho}\overline{\theta}\) is the mass-weighted potential temperature of the reference atmosphere. Consequently, \cref{eq:large-scale-eq-full} can be rewritten as 
\begin{equation}
    \label{eq:large-scale-eq-leading}
    \begin{split}
        \varepsilon\left(\partial_{\ct} 
            + 
            \mean{\underline{\vb{V}}}\cdot\nabla_{\cxv}
        \right)
        \mean{\underline{\Psi}} 
            = 
            \mathcal{Q}^{(0)} + \mathcal{Q}^{(1)} + O\left(\varepsilon^{4}\right) \;.
    \end{split}
\end{equation}
The leading-order terms on the right-hand side of \cref{eq:large-scale-eq-full}, summarized in \cref{eq:large-scale-eq-leading} as \(\mathcal{Q}^{(0)}\), are \(\mathcal{O}(\varepsilon)\) smaller than the large-scale advection and can be rewritten as 
\begin{equation}
    \label{eq:leading-order-gw-forcing}
    \mathcal{Q}^{(0)}\equiv -\frac{\varepsilon^{2}}{2\overline{P}}\nabla_{\cxv}\cdot\mathfrak{R}\left(\overline{P}\hat{\vb{V}}^{(0)}\hat{\Psi}^{(1)^*}\right) \;.
\end{equation}
Similarly, all terms in \cref{eq:large-scale-eq-full} that are \(\mathcal{O}\left(\varepsilon^2\right)\) smaller than the large-scale advection are grouped into \(\mathcal{Q}^{(1)}\) in \cref{eq:large-scale-eq-leading} and can be rewritten as 
\begin{equation}
    \label{eq:next-order-gw-forcing}
    \begin{split}
        \mathcal{Q}^{(1)} \equiv -\frac{\varepsilon^{3}}{2\overline{P}}&\nabla_{\cxv}\cdot\mathfrak{R}\left[\overline{P}\left(\hat{\vb{V}}^{(0)}\hat{\Psi}^{(2)^*} + \hat{\vb{V}}^{(1)}\hat{\Psi}^{(1)^*}\right)\right]  \;.
    \end{split}
\end{equation}
The term \(\mathcal{Q}^{(0)}\) corresponds to tracer transport by the Stokes drift associated with gravity waves. In contrast, \(\mathcal{Q}^{(1)}\) captures higher-order nonlinear contributions, which originate from secondary wave fields generated by the primary (first-order) waves.

With gravity wave parameterization in mind, it is convenient to express the terms in \cref{eq:leading-order-gw-forcing,eq:next-order-gw-forcing} using variables commonly computed in parameterization models, such as the wave action density and wavenumbers. This formulation will be presented in the following sections.

\subsection{Leading-order gravity wave tracer flux convergence}

In this section, we describe how to calculate the leading-order gravity wave tracer flux convergence, \(\mathcal{Q}^{(0)}\), in \cref{eq:leading-order-gw-forcing} using the wave action density and wavenumbers predicted by a gravity wave model.

We aim to describe \(\mathcal{Q}^{(0)}\) in terms of the wave action density, \(\mathcal{A} = E_{w}/\hat{\omega}\), where \(E_{w,}\) is the wave energy given by  \citet{Achatz2023MultiScaleDynamics} as
\begin{equation}
    \label{eq:wave-energy}
    E_{w} = \frac{\overline{\rho}}{2}
    \left(\frac{\left|\hat{\vb{U}}^{(0)}\right|^2}{2} 
    + \varepsilon^{4}\frac{\left|\hat{W}^{(0)}\right|^2}{2}  
    + \frac{1}{N_0^2}\frac{\left|\hat{B}^{(0)}\right|^2}{2}
    \right) 
    = \overline{\rho} \frac{\left|\hat{B}^{(0)}\right|^2}{2 N_0^2} 
    \frac{N_0^2k_{h}^2+f_0^2m^2}{N_0^2 k_{h}^2} \;,
\end{equation}
\(\hat{B}^{(0)}=\hat{\Theta}^{(0)}/\overline{\underline{\Theta}}\) is the leading-order buoyancy amplitude, and the horizontal wavenumber \(k_{h}\) is defined as \(k_{h}=\sqrt{k^2+l^2}\). 
The dispersion relation for internal gravity waves is given by 
\begin{equation}
    \label{eq:dispersion-relation}
    \hat{\omega}^2 = \frac{N_0^2k_{h}^2+f_0^2m^2}{\varepsilon^{4}k_{h}^2+m^2} \;.
\end{equation}
Therefore, we must express the leading-order tracer and wind amplitudes, \(\hat{\Psi}^{(1)}\) and \(\hat{\vb{V}}^{(0)}\), respectively, in terms of the buoyancy amplitude \(\hat{B}^{(0)}\). 
The tracer amplitude is coupled to the wind amplitude via \cref{eq:amplitude-eq-full-gamma0}. The wind amplitude, in turn, is couple to the buoyancy amplitude through the polarization relations  \citep{Achatz2017Interaction,Achatz2023MultiScaleDynamics} 
\begin{align}
    \label{eq:polarization-u}
    \hat{\vb{U}}^{(0)} &= \frac{i}{mN_0^2}\frac{\varepsilon^{4}\hat{\omega}^2-N_0^2}{\hat{\omega}^2-f_0^2}\left(\vb{k}_{h}\hat{\omega}-if_0\vb{e}_z\times\vb{k}_{h}\right)\hat{B}^{(0)}\;, \\
    \label{eq:polarization-w}
    \hat{W}^{(0)} &= \frac{i\hat{\omega}}{N_0^2}\hat{B}^{(0)} \;, \\
    \label{eq:polarization-pi}
    \frac{c_p}{R}\underline{\overline{\Theta}}\hat{\Pi}^{(0)} &=
    \frac{i}{m}\frac{\varepsilon^{4}\hat{\omega}^2-N_0^2}{N_0^2}
    \hat{B}^{(0)} \;.
\end{align}
As it is needed later on, we have also included the polarization relation for the Exner pressure. 
These relations are derived from the matrix equation 
\begin{equation} \label{eq:matrix-equation-0}
    M\vb{Z}^{(0)} = 0 \;,
\end{equation}
where the matrix \(M\) is given by
\begin{equation}
    \label{eq:matrix}
    M=M(\vb{k},\hat{\omega}) = 
    \mqty[-i\hat{\omega} & -f_0 & 0 & 0 & i k \\
    f_0 & -i\hat{\omega} & 0 & 0 & i l \\
    0 & 0 & -i\hat{\omega}\varepsilon^{4} & -N_0 & i m \\
    0 & 0 & N_0 & -i\hat{\omega} & 0 \\
    i k & i l & i m & 0 & 0] \;,
\end{equation}
the components of the wave numbers are denoted by \(\vb{k} = (k,l,m)^{\mathrm{T}}\), 
and the vector \(\vb{Z}^{(0)}\) contains the wave amplitudes
\begin{equation}
   \vb{Z}^{(0)} = \left(
    \hat{U}^{(0)}\:,\:
    \hat{V}^{(0)}\:,\:
    \hat{W}^{(0)}\:,\:
    \hat{B}^{(0)}/N_0\:,\:
    \frac{c_p}{R}
    \overline{\underline{\Theta}}
    \hat{\Pi}^{(0)}\right)^{\mathrm{T}} \;.
\end{equation}
The non-dimensional Brunt-Väisälä frequency is defined as \(N_0^2=\dd_{\cz}\overline{\Theta}^{(1)}/\overline{\underline{\Theta}}\).

By inserting \cref{eq:amplitude-eq-full-gamma0,eq:polarization-u,eq:polarization-w,eq:dispersion-relation,eq:wave-energy} into \cref{eq:leading-order-gw-forcing}, we obtain an expression for \(\mathcal{Q}^{(0)}\) in terms of the wave action density:
\begin{equation}
    \mathcal{Q}^{(0)} = - \frac{\varepsilon}{\overline{\rho}}\nabla_{\cxv}\cdot
    \left(
        \frac{f_0}{\hat{\omega}}\frac{m}{\varepsilon^{4}k_{h}^2+m^2}\mathcal{A}
    \vb{k}\times\nabla_{\cxv}\mean{\underline{\Psi}}
    \right) 
    \;.
\end{equation}

\subsection{Next-order gravity wave tracer flux convergence}

As shown in the previous section, the leading-order tracer flux convergence \(\mathcal{Q}^{(0)}\) arises from inertia-gravity waves and is the dominant gravity-wave forcing on tracers in rotating flows. Because the Coriolis frequency decreases toward  the equator, this leading-order contribution weakens. In equatorial settings, such as those influenced by tropical convective gravity waves, the tracer forcing must come from higher-order terms. Although the global impact of the next-order convergence \(\mathcal{Q}^{(1)}\) is likely small, it becomes relevant in idealized or weak-rotation regimes, including simulations without planetary rotation. For completeness and to enable accurate tests of parameterized waves under such conditions, we therefore derive a method to determine \(\mathcal{Q}^{(1)}\) from the wave action density and wavenumbers.

The next-order flux convergence, \(\mathcal{Q}^{(1)}\), is given in \cref{eq:next-order-gw-forcing}. 
Since \(\mathcal{A}\propto\left|\hat{B}^{(0)}\right|^2\), both the leading-order and next-order tracer and wind amplitudes must be expressed in terms of the buoyancy amplitude. 
As described in the previous section, the leading-order amplitudes are coupled to \(\hat{B}^{(0)}\) via \cref{eq:amplitude-eq-full-gamma0,eq:polarization-u,eq:polarization-w}. 
The next-order tracer amplitudes can be obtained from the next-order wind amplitudes and the leading-order tracer amplitudes, as shown in \cref{eq:amplitude-eq-full-gamma1}. 
Therefore, the remaining task is to determine \(\hat{\vb{V}}^{(1)}\) in terms of \(\hat{B}^{(0)}\).
Similar to the leading-order basic wave amplitudes, the next-order wind, buoyancy and Exner pressure amplitudes \(\hat{\vb{V}}^{(1)}\), \(\hat{B}^{(1)}=\hat{\Theta}^{(1)}/\overline{\underline{\Theta}}\), and \(\hat{\Pi}^{(1)}\), respectively, satisfy a matrix equation 
\begin{equation}
    \label{eq:matrix-eq-1}
    M\vb{Z}^{(1)} = \vb{R}^{(1)} \;,
\end{equation}
where \(M\) given in \cref{eq:matrix}, 
\begin{equation}
    \vb{Z}^{(1)} = \left(
    \hat{U}^{(1)}\:,\:
    \hat{V}^{(1)}\:,\:
    \hat{W}^{(1)}\:,\:
    \hat{B}^{(1)}/N_0\:,\:
    \frac{c_p}{R}
    \overline{\underline{\Theta}}
    \hat{\Pi}^{(1)}\right)^{\mathrm{T}} \;,
\end{equation}
and
\begin{equation}
    \vb{R}^{(1)} = \left(
    R_{u}\:,\:
    R_{v}\:,\:
    R_{w}\:,\:
    R_{b}/N_0\:,\:
    R_{\pi}\right)^{\mathrm{T}} \;.
\end{equation}
The components of \(\vb{R}^{(1)}\) are 
\begin{align}
    \begin{split}
        \vb{R}_{\vb{u}}^{(1)} =& - \left(\partial_{\ct} + \mean{\underline{\vb{U}}}\cdot\nabla_{\cxv,h}\right)\hat{\vb{U}}^{(0)} 
        - \left(\hat{\vb{V}}^{(0)}\cdot\nabla_{\cxv}\right)\mean{\underline{\vb{U}}} 
        - \frac{c_p}{R}\left[
        \overline{\underline{\Theta}} \nabla_{\cxv,h}\hat{\Pi}^{(0)} + \varepsilon
        \left(i\vb{k}_{h}\mean{\underline{\Theta}}
        \hat{\Pi}^{(0)}+ \hat{\Theta}^{(0)}\nabla_{\cxv,h}\mean{\underline{\Pi}} \right)\right] \;, 
    \end{split} \\
    \begin{split}
        R_{w}^{(1)} =& -\varepsilon^{4}\left(\partial_{\ct} + \mean{\underline{\vb{U}}}\cdot\nabla_{\cxv,h}\right)\hat{W}^{(0)} 
        - \frac{c_p}{R}\left[\overline{\underline{\Theta}} \partial_{\cz}\hat{\Pi}^{(0)} + 
        \varepsilon
        \left(im\mean{\underline{\Theta}}\hat{\Pi}^{(0)} + \hat{\Theta}^{(0)} 
        \partial_{\cz}\mean{\underline{\Pi}}\right)\right] \;,
    \end{split} \\
    \begin{split}
        R_{b}^{(1)} =& -\left(\partial_{\ct} + \mean{\underline{\vb{U}}}\cdot\nabla_{\cxv,h}\right)
        \hat{B}^{(0)}
        - \frac{\hat{\vb{V}}^{(0)}}{\overline{\underline{\Theta}}}
        \cdot\nabla_{\cxv}\mean{\underline{\Theta}} \;,
    \end{split} \\
    \begin{split}
        R_{\pi}^{(1)} 
        =& 
        - \frac{1}{\overline{P}}
        \nabla_{\cxv}\cdot
        \left(
            \overline{P}
            \hat{\vb{V}}^{(0)}
        \right) \;,
    \end{split}
\end{align}
with \(\vb{R}_{\vb{u}}^{(1)} = \left(R_{u}^{(1)}\:,\:R_{v}^{(1)}\right)^{\mathrm{T}}\).
These right-hand sides follow \citet{Achatz2017Interaction}. 
However, in contrast to that work, we have included higher-order terms to facilitate a unified treatment of the two stratification cases (see \cref{sec:appendix-full-expansion}). Likewise, we retain the full expansion of the advecting large-scale winds. 
Using the polarization relations from \cref{eq:polarization-u,eq:polarization-w,eq:polarization-pi}, all right-hand sides can be determined form the spatial and temporal dependence of the wavenumber and leading-order buoyancy amplitude \(\hat{B}^{(0)}\). Once these terms are known, the next-order wave amplitudes \(\vb{Z}^{(1)}\) can be obtained by solving \cref{eq:matrix-eq-1}.

The leading-order buoyancy amplitude, required for all subsequent calculation, can be decomposed into its absolute magnitude and phase factor 
\begin{equation}
    \label{eq:buoyancy-decomp}
    \hat{B}^{(0)} = \left|\hat{B}^{(0)}\right|e^{i\varLambda} \;,
\end{equation}
where \(\varLambda\) represents the large-scale wave phase. The absolute magnitude can be obtained from the wave energy \(E_{w}\) via \cref{eq:wave-energy}. While \(E_{w}\) is directly available from the gravity wave model, the large-scale phase must be predicted from the prognostic equation 
\begin{equation}
    \label{eq:wave-phase-equation}
    \begin{split}
        \left(\partial_{\ct} + \vb{c}_{g}\cdot\nabla_{\cxv}\right)\varLambda
        =& - \frac{f_0m}{2\left|\vb{k}\right|^2\hat{\omega}}
        \left(\vb{k}\times\nabla_{\cxv}\right)\mean{\underline{\vb{U}}} \\
        &+ \frac{|\vb{k}|^2}{2m\left(N_0^2-\varepsilon^4f_0^2\right)}
        \left[\frac{1}{\overline{\underline{\Theta}}}\left(\hat{\vb{c}}_g\cdot\nabla_{\cxv}\right)\mean{\underline{\Theta}} - \varepsilon N_0^2
        \frac{c_p}{R}\overline{\underline{\Theta}}
        \left(\hat{\vb{c}}_g\cdot\nabla_{\cxv}\right)\mean{\underline{\Pi}}\right] \\
        &- \frac{f_0\hat{\omega}m^2}{k_h^2\left(N_0^2-\varepsilon^4f_0^2\right)}
        \left[\left(l\partial_{\cx}-k\partial_{\cy}\right)\hat{\omega}+l\left(\partial_{\ct}+\mean{\underline{\vb{U}}}\cdot\nabla_{\cxv,h}\right)k
        - k\left(\partial_{\ct}+\mean{\underline{\vb{U}}}\cdot\nabla_{\cxv,h}\right)l\right]
    \end{split}
\end{equation}
where \(\vb{c}_{g}=\nabla_{\vb{k}}\omega\) is the group velocity and \(\hat{\vb{c}}_{g}=\nabla_{\vb{k}}\hat{\omega}\) is the intrinsic group velocity.
We derive the prognostic equation by taking the imaginary part of the solvability condition, as outlined in \citet{Achatz2010Gravity}:
\begin{equation}
    0 = \vb{Z}^{(0)^\dagger}\vb{R}^{(1)} \;.
\end{equation}
In summary, \(\hat{B}^{(0)}\) in \cref{eq:buoyancy-decomp} can be calculated using \cref{eq:wave-energy,eq:wave-phase-equation}. 
The leading-order tracer and wind amplitudes can then be determined from \cref{eq:amplitude-eq-full-gamma0} and \cref{eq:polarization-u,eq:polarization-w}, respectively. 
Next, by solving \cref{eq:matrix-eq-1} for \(\vb{Z}^{(1)}\) and using \cref{eq:amplitude-eq-full-gamma1}, the next-order tracer and wind amplitudes can be computed. Finally, the resulting \(\hat{\Psi}^{(1)}\), can be substituted into \cref{eq:next-order-gw-forcing} to obtain \(\mathcal{Q}^{(1)}\).

\section{Summary of equations in dimensional form} \label{sec:summary-equations}

Finally, we summarize the most important equations in dimensional form. To achieve this, we make the following substitutions: 
\begin{align}
    \hat{\omega} &\rightarrow \hat{\omega}\widetilde{T} = \hat{\omega}/f \\
    (\cx,\cy,\cz) &\rightarrow \varepsilon\left[(x,y)/\widetilde{L},z/\widetilde{H}\right] \\
    \ct &\rightarrow \varepsilon t/\widetilde{T} \\
    (k,l,m) &\rightarrow \left[\widetilde{L}(k,l),\widetilde{H}m\right] \\
    f_0 &\rightarrow f/f \\
    \overline{\underline{\Theta}} &\rightarrow \overline{\theta}/T_{00} \\
    \left[\mean{\underline{\vb{U}}},\:\mean{\underline{W}},\:\mean{\underline{\Psi}},\:
    \mean{\underline{\Theta}},\:\mean{\underline{\Pi}}\right] &\rightarrow 
    \left[\frac{\mean{\vb{u}}}{\widetilde{U}},\:
    \frac{\mean{w}}{\varepsilon\widetilde{W}},\:
    \frac{\mean{\psi}}{\widetilde{\Psi}},\:
    \frac{\mean{\theta}}{\varepsilon^{2}T_{00}},\:
    \frac{\mean{\pi}}{\varepsilon^{2}}\right] \;.
\end{align}
With the dispersion relation 
\begin{equation} \label{eq:dispersion_relation}
    \hat{\omega}^2 = \frac{N^2k_{h}^2+f^2m^2}{k_{h}^2+m^2} \;,
\end{equation}
where \(N^2 = (g/\overline{\theta})\dd_z\overline{\theta}\), and the substitution for the leading-order wave amplitudes 
\begin{equation}
    \left(\hat{\vb{U}}^{(0)}\:,\:
    \hat{W}^{(0)}\:,\:
    \hat{B}^{(0)}\:,\:
    \hat{\Pi}^{(0)}\:,\:
    \hat{\Psi}^{(1)}
    \right) \rightarrow 
    \left(
    \frac{\vb{u}'}{\widetilde{U}}\:,\:
    \frac{w'}{\widetilde{W}}\:,\:
    \frac{\theta'}{\varepsilon^{2}\overline{\theta}}\:,\:
    \frac{\pi'}{\varepsilon^{3}}\:,\:
    \frac{\psi'}{\varepsilon\widetilde{\Psi}}
    \right) \;,
\end{equation}
we obtain the polarization relations 
\begin{align}
    \label{eq:polarization-u-dim}
    \vb{u}' &= \frac{i}{mN^2}\frac{\hat{\omega}^2-N^2}{\hat{\omega}^2-f^2}\left(\vb{k}_{h}\hat{\omega}-if\vb{e}_z\times\vb{k}_{h}\right)b' \;, \\
    \label{eq:polarization-w-dim}
    w' &= \frac{i\hat{\omega}}{N^2}b' \;, \\
    \label{eq:polarization-pi-dim}
    c_p\overline{\theta}\pi' &= \frac{i}{m}\frac{\hat{\omega}^2-N^2}{N^2}b' \;.
\end{align}
The leading-order tracer amplitude is given by 
\begin{align}
    \label{eq:tracer_lead_dim}
    \psi' &= -\frac{i}{\hat{\omega}}\vb{v}'\cdot\nabla\mean{\psi} \;.
\end{align}
The re-dimensionalized matrix equation~(\ref{eq:matrix-eq-1}) is expressed as 
\begin{equation}
    \label{eq:matrix-eq-1-redim}
    \underbrace{\mqty[-i\hat{\omega} & -f & 0 & 0 & ik \\
    f & -i\hat{\omega} & 0 & 0 & il \\
    0 & 0 & -i\hat{\omega} & -N & im \\ 
    0 & 0 & N & -i\hat{\omega} & 0 \\
    ik & il & im & 0 & 0]}_{M}
    \underbrace{\mqty[u'' \\ v'' \\ w'' \\ 
    b''/N \\ c_p\overline{\theta}\pi'']}_{\vb{Z}''} = 
    \underbrace{\mqty[R_{u} \\ R_{v} \\ R_{w} \\
    R_{b}/N \\ R_{\pi}]}_{\vb{R}} \;,
\end{equation}
where the next-order wave amplitudes are obtained through the substitution 
\begin{equation}
    \left(\hat{\vb{U}}^{(1)}\:,\:
    \hat{W}^{(1)}\:,\:
    \hat{B}^{(1)}\:,\:
    \hat{\Pi}^{(1)}\:,\:
    \hat{\Psi}^{(2)}
    \right) \rightarrow 
    \left(
    \frac{\vb{u}''}{\widetilde{U}}\:,\:
    \frac{w''}{\widetilde{W}}\:,\:
    \frac{\theta''}{\varepsilon^{2}\overline{\theta}}\:,\:
    \frac{\pi''}{\varepsilon^{3}}\:,\:
    \frac{\psi''}{\varepsilon\widetilde{\Psi}}
    \right)\frac{1}{\varepsilon} \;,
\end{equation}
and the components of \(\vb{R}\) are 
\begin{align}
    \label{eq:rhs_u_dim}
    \begin{split}
        \vb{R}_{\vb{u}} =& -\left(\partial_t + \mean{\vb{u}}\cdot\nabla_h 
        \right)\vb{u}' 
        - \vb{v}'
        \cdot\nabla\mean{\vb{u}}  
        - c_p
        \left[
            \overline{\theta}\nabla_h
            \pi'
            + i\vb{k}_{h}\mean{\theta}\pi'+\theta'\nabla_h\mean{\pi}
        \right]\;,
    \end{split} \\
    \label{eq:rhs_w_dim}
    \begin{split}
        R_{w} =& -\left(\partial_t + \mean{\vb{u}}\cdot\nabla_h
        \right)w' - c_p
        \left[
            \overline{\theta}\partial_z\pi'  
        + im\mean{\theta}\pi' + \theta'\partial_z\mean{\pi}
        \right]\;,
    \end{split} \\
    \label{eq:rhs_b_dim}
    \begin{split}
        R_{b} 
        =& 
        -\left(\partial_t + \mean{\vb{u}}\cdot\nabla_h 
        \right)
        b' 
        - \frac{\vb{v}'}{\overline{\theta}}
        \cdot\nabla\mean{\theta} \;,
    \end{split} \\
    \label{eq:rhs_pi_dim}
    \begin{split}
        R_{\pi} 
        =& -\frac{1}{
            \overline{\rho}
        \overline{\theta}}
        \nabla\cdot
        \left(
            \overline{\rho}
            \overline{\theta}
            \vb{v}'
        \right) \;.
    \end{split}
\end{align}
The next-order tracer amplitude is given by 
\begin{equation}
    \label{eq:tracer_next_dim}
    i\hat{\omega}
    \psi'' 
    = 
    -\left(\partial_t + \mean{\vb{u}}\cdot\nabla_h 
    \right)
    \psi' 
    + \vb{v}''
    \cdot\nabla\mean{\psi} \;.
\end{equation}
Re-dimensionalized, the tracer flux convergence due to the Stokes drift \(\mathcal{Q}^{(0)}\) is
\begin{equation} \label{eq:leading-order-flux-convergence-redim}
    \mathcal{Q}^{(0)} = -\frac{1}{\overline{\rho}}\nabla\cdot\frac{f}{\hat{\omega}}\frac{m}{k_{h}^2+m^2}\mathcal{A}
    \vb{k}\times\nabla\mean{\psi}
    \;,
\end{equation}
representing the convergence of tracer fluxes due to the inertia-gravity wave Stokes drift from the gradient of the large-scale tracer. The next-order gravity wave-impact
\begin{equation}
    \label{eq:next-order-gw-forcing_dim}
    \begin{split}
        \mathcal{Q}^{(1)} 
        = 
        -\frac{1}{2 \overline{\rho}\overline{\theta}}
        &\nabla\cdot\mathfrak{R}
        \left[
            \overline{\rho}\overline{\theta}
            \left(
                \vb{v}'\psi''^* + \vb{v}''\psi'^*
                \right)\right] \;,
    \end{split}
\end{equation}
arising from nonlinear effects associated with next-order wave fields \(\vb{v}''\) and \(\psi''\), generated by the first-order fields (\cref{eq:matrix-eq-1-redim,eq:tracer_next_dim}). \(\mathcal{Q}^{(1)}\) is determined through a sequential process, under the assumption that the wave energy \(E_w\) is known:
\begin{enumerate}
    \item Solving for the leading-order buoyancy amplitude \(b'=\left|b'\right|e^{i\varLambda}\) from the wave energy equation 
    \begin{equation}
        E_{w} 
        = \frac{\overline{\rho}}{2}\left(\frac{\left|\vb{v}'\right|^2}{2} + \frac{\left|b'\right|^2}{2N^2} \right)
        = \overline{\rho} \frac{\left|b'\right|^2}{2 N^2} \frac{N^2k_{h}^2+f^2m^2}{N^2 k_{h}^2} \;.
    \end{equation}
    Note that this relation provides \(|b'|\) once the wave energy \(E_w\) is specified; in practice \(E_w\) must be predicted by a gravity-wave model or another energy source.
    \item Integrating the prognostic equation for \(\varLambda\) to obtain its large-scale phase:
    \begin{equation}
        \label{eq:ls_phase_dim}
        \begin{split}
            \left(\partial_t + \vb{c}_{g}\cdot\nabla\right)\varLambda 
            =&
            - \frac{fm}{2\left|\vb{k}\right|^2\hat{\omega}}\left(\vb{k}\times\nabla\right)\mean{\vb{u}} \\
            &+ \frac{|\vb{k}|^2}{2m\left(N^2-f^2\right)}
            \left[\frac{g}{\overline{\theta}}\left(\hat{\vb{c}}_{g}\cdot\nabla\right)\mean{\theta}
            -\frac{N^2}{g}c_p\overline{\theta}\left(\hat{\vb{c}}_{g}\cdot\nabla\right)\mean{\pi}\right] \\
            &- \frac{f\hat{\omega}m^2}{k_h^2\left(N^2k_h^2+f^2m^2\right)}
            \left[\left(l\partial_x+k\partial_y\right)\hat{\omega}
            + l\left(\partial_t+\mean{\vb{u}}\cdot\nabla_h\right)k
            - k\left(\partial_t+\mean{\vb{u}}\cdot\nabla_h\right)l\right] \;.
        \end{split}
    \end{equation}
    \item Using the polarization relations (\cref{eq:polarization-u-dim,eq:polarization-w-dim,eq:polarization-pi-dim}) to compute the leading-order wind and pressure amplitudes.
    \item Determining the right-hand sides (\cref{eq:rhs_u_dim,eq:rhs_w_dim,eq:rhs_b_dim,eq:rhs_pi_dim} of \cref{eq:matrix-eq-1-redim}) of \cref{eq:matrix-eq-1-redim} based on the leading-order wave amplitudes.
    \item Solving \cref{eq:matrix-eq-1-redim} for the next-order wind amplitudes.
    \item Computing the leading-order tracer amplitude from \cref{eq:tracer_lead_dim} and the next-order tracer amplitude from \cref{eq:tracer_next_dim}.
    \item Finally, substituting these values, along with the leading-order and next-order wind amplitudes, into \cref{eq:next-order-gw-forcing_dim} to obtain \(\mathcal{Q}^{(1)}\).
\end{enumerate}
A difficulty arises in step 5: As the dispersion relation \(\hat{\omega}\) (\cref{eq:dispersion_relation}) is obtained by enforcing \(\det(M)=0\), the inverse of \(M\) cannot be computed directly. However, it is important to note that the next-order amplitude vector \(\vb{Z}''\) does not belong to the null space of \(M\), as the latter is assumed to only contribute to the leading-order wave amplitudes \(\vb{Z}'\) (\cref{eq:matrix-equation-0}).  This allows us to determine the exact solution for \(\vb{Z}''\) by determining the inverse using the singular value decomposition \citep[SVD;][]{Golub1965}. Writing
\begin{equation}
    M = U\Sigma V^* \;,
\end{equation}
with unitary matrices \(U\) and \(V\), and diagonal \(\Sigma\) containing the singular values, the pseudo-inverse is 
\begin{equation}
    M^{-1} = V \Sigma^{-1} U^* \;,
\end{equation}
where \(\Sigma^{-1}\) is obtained by inverting non-zero singular values. This yields the exact solution 
\begin{equation}
    \vb{Z}'' = M^{-1} \vb{R} \;.
\end{equation}

\section{Numerical Simulation of Gravity Wave-Induced Tracer Transport} \label{sec:numerical-model}

We extend a gravity wave parameterization model to compute the previously discussed gravity wave tracer flux convergences. The model is coupled with a flow solver, both of which are introduced in \cref{sec:numerical-model2}. The treatment of leading-order and next-order flux convergences is detailed in \cref{sec:leading-order-results,sec:next-order-results}, respectively. To validate our approach, we compare parameterized gravity wave simulations at coarse resolution against high-resolution, wave-resolving reference simulations.

\subsection{Numerical Model} \label{sec:numerical-model2}

In the following section, we introduce the Pseudo-Incompressible Flow solver (PincFlow) used to simulate the resolved flow. PincFlow operates in two modes: a high-resolution, wave-resolving mode for generating reference data and a low-resolution mode with parameterized gravity waves, incorporating a Lagrangian ray tracer. We then present the gravity wave model, which computes the gravity wave impact on the resolved flow and is coupled to PincFlow.

\subsubsection{PincFlow}

PincFlow solves the pseudo-incompressible equations, first introduced by \citet{Durran1989Improving}, in conservative flux form, as presented by \citet{Klein2009Asymptotics}, \citet{Rieper2013ConservativeIntegration}, \citet{Schmid2021NumericalLaboratory}, and \citet{Jochum2025}. The model includes a passive tracer mixing ratio, \(\psi\), whose transport is described by \cref{eq:tracer-eq-dim-form} and, in flux form, by
\begin{equation} \label{eq:tracer-advection-flux}
    \pdv{(\rho\psi)}{t} + \nabla\cdot(\rho\psi\vb{v}) = 0\;.
\end{equation}
Atmospheric variables are discretized on a C-grid \citep{ARAKAWA1977}, where density, Exner pressure, and tracer mixing ratio are defined at cell centers, while the wind components \(u\), \(v\), and \(w\) are staggered along the \(x\)-, \(y\)-, and \(z\)-directions, respectively. Mass, tracer, and momentum fluxes are computed using a monotone upwind scheme \citep{Leer1979Upwind}. The equations are integrated using a semi-implicit time-stepping scheme \citep{Schmid2021NumericalLaboratory}. The tracer advection equation (\cref{eq:tracer-advection-flux}) is solved explicitly twice per time step, each over half a step. For explicit time integration, the third-order Runge-Kutta method of \citet{Williamson1980} is employed.

\subsubsection{PincFlow/MS-GWaM} \label{sec:msgwam}

For coarse-resolution simulations using PincFlow, the gravity wave impact on the large-scale flow can be parameterized with the Multi-Scale Gravity-Wave Model (MS-GWaM). The modeled gravity waves are represented by a spectral extension of the WKB theory \citep{Achatz2023MultiScaleDynamics} and integrates it using a Lagrangian approach. Below, we provide a brief overview of MS-GWaM, while more detailed descriptions can be found in \citet{Muraschko2015Application}, \citet{Boloni2016Interaction}, \citet{Wilhelm2018Interaction}, \citet{Wei2019Efficient}, and \citet{Jochum2025}.

MS-GWaM calculates the propagation of phase-space (or spectral) wave action density \(\mathcal{N}\), where
\begin{equation}
    \mathcal{N}(\vb{x},\vb{k},t) = \sum_{\alpha=1}^{M}\mathcal{A}_{\alpha}(\vb{x},t)\var{[\vb{k}-\vb{k}_{\alpha}(\vb{x},t)]} \;,
\end{equation}
using the conservation equation
\begin{equation}
    \label{eq:wave-action-conservation}
    \dv{\mathcal{N}}{t} \equiv \pdv{\mathcal{N}}{t} + \vb{c}_g\cdot\nabla_{\vb{x}}\mathcal{N} + \dot{\vb{k}}\cdot\nabla_{\vb{k}}\mathcal{N} = 0 \;,
\end{equation}
meaning \(\mathcal{N}\) is conserved along ray trajectories. These trajectories are defined by the group velocity and rate of change of wavenumber
\begin{align}
    \dv{\vb{x}}{t} &= \vb{c}_g 
    = \nabla_{\vb{k}}\Omega \;, \\ 
    \dv{\vb{k}}{t} &= \dot{\vb{k}} 
    = -\nabla_{\vb{x}}\Omega\;,
\end{align}
respectively, where
\begin{equation}
    \Omega(\vb{x},\vb{k},t) 
    = 
    \vb{k}_h\cdot\langle\vb{u}\rangle(\vb{x},t)
    \pm
    \sqrt{\frac{k_h^2 N^2(z) + f^2 m^2}{k_h^2 + m^2}}
\end{equation}
represents the gravity-wave dispersion relation.

Numerically, the problem is discretized by dividing phase-space regions with nonzero wave action density into six-dimensional cuboid ray volumes, extending in the \(x\)-, \(y\)-, \(z\)-, \(k\)-, \(l\)-, and \(m\)-directions. Since the phase-space velocity is divergence-free: 
\begin{equation}
    \nabla_{\vb{x}}\cdot\vb{c}_g + \nabla_{\vb{k}}\cdot\dot{\vb{k}} = 0 \;,
\end{equation}
the volume content of each ray remains constant during propagation. While ray volumes maintain their cuboid shape—meaning shear effects are only captured by subdividing sheared regions—they can stretch or compress in a volume-preserving manner.

In the next-two sections, we extend MS-GWaM to calculate the leading- and next-order gravity wave tracer flux convergences, as discussed in \cref{sec:theory,sec:summary-equations}. To validate these extensions, we use two test cases of gravity wave packets, each designed so that one of the flux convergences dominates the gravity wave forcing on the large-scale tracer.

\subsection{Parameterization of the Leading-order Gravity Wave Tracer Flux Convergence} \label{sec:leading-order-results}

\begin{figure}[t]
 \noindent\includegraphics[width=\textwidth,angle=0]{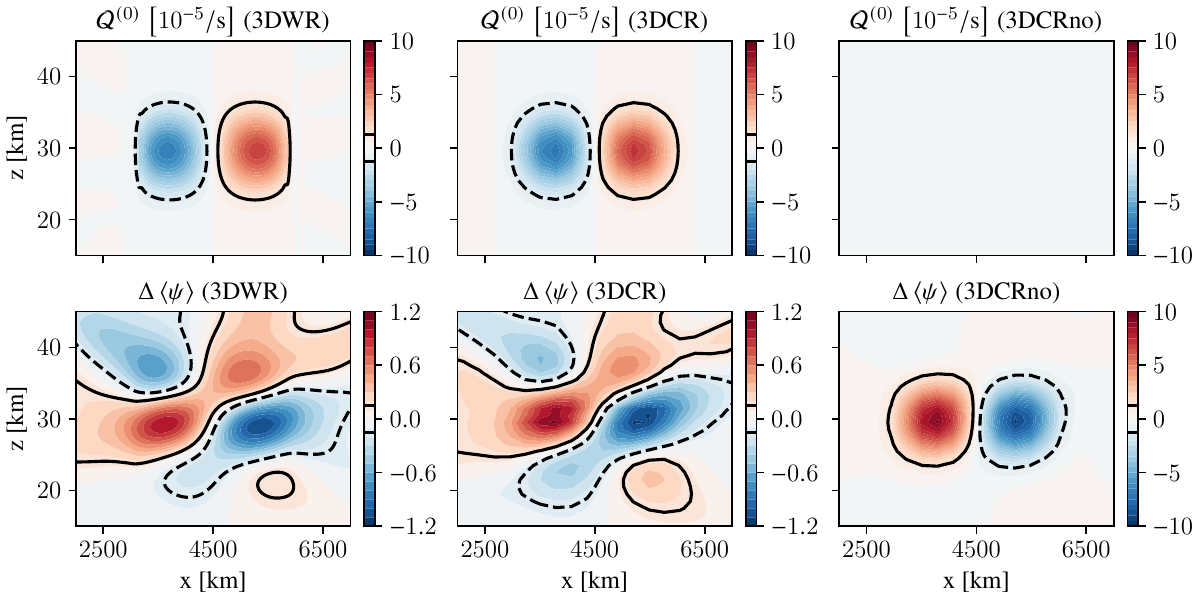}
 \caption{Gravity wave-induced tracer flux convergence \(\mathcal{Q}^{(0)}\) (top row) and change in the large-scale tracer distribution \(\Delta\mean{\psi}\) (bottom row) from a three-dimensional gravity wave packet in an atmosphere initially at rest, at \(t=1500\text{min}\); (left column) 3DWR: wave-resolving reference simulation; (center column) 3DCR: coarse-resolution simulation with parameterized gravity wave and including \(\mathcal{Q}^{(0)}\); (right column) 3DCRno: coarse-resolution simulation with \(\mathcal{Q}^{(0)}=\mathcal{Q}^{(1)}=0\). The solid and dashed contour lines illustrate the regions with positive and negative values, respectively.}\label{fig:2DWP90_comparison}
\end{figure}

In this section, we present the parameterization of the leading-order gravity wave tracer flux convergence using MS-GWaM. To validate this extension, we conduct two simulations of three-dimensional gravity wave packets propagating through a large-scale tracer field that initially varies only in the vertical direction. We then compare the resulting changes in the large-scale tracer distribution with those from a wave-resolving reference simulation.

To adapt the theory described above to the spectral Lagrangian approach of MS-GWaM, we rewrite the leading-order flux convergence in terms of \(\mathcal{N}\). From \cref{eq:leading-order-flux-convergence-redim}, we obtain 
\begin{equation}
    \label{eq:leading-order-fluxes-msgwam}
    \mathcal{Q}^{(0)} = - \frac{1}{\overline{\rho}}\nabla\cdot\int\dd[3]{k}\frac{f}{\hat{\omega}}\frac{m\mathcal{N}}{ k_h^2+m^2}\left(\vb{k}\times\nabla\psi\right) \;.
\end{equation} 
It is important to note that the derivation of \cref{eq:wave-action-conservation} assumes geostrophic and hydrostatic equilibrium of the large-scale flow to ensure the numerical feasibility of simulation \(\mathcal{N}\), as presented in \citet{Achatz2023MultiScaleDynamics}. However, this assumption is not required for the derivation of \(\mathcal{Q}^{(0)}\) or \(\mathcal{Q}^{(1)}\). The geostrophic and hydrostatic equilibrium approximation introduces an \(\mathcal{O}(\varepsilon)\) error in the prognostic equation~(\ref{eq:wave-action-conservation}) for \(\mathcal{N}\). As demonstrated by \citet{Boloni2016Interaction}, \citet{Wei2019Efficient}, and \citet{Jochum2025}, and as will be confirmed by the following simulation results, this error appears to be sufficiently small for the predicted \(\mathcal{N}\) to yield reliable results.

The impact of gravity waves on tracer transport is incorporated by extending \cref{eq:tracer-advection-flux} to 
\begin{equation} \label{eq:tracer-advection-flux-Q0}
    \pdv{(\rho\psi)}{t} + \nabla\cdot(\rho\psi\vb{v}) = \rho \mathcal{Q}^{(0)}\;.
\end{equation}
While \cref{eq:tracer-advection-flux} is used in wave-resolving reference simulations, the extended form, \cref{eq:tracer-advection-flux-Q0}, is applied when the model resolution is coarse and gravity waves are parameterized using MS-GWaM.

To validate the extension of PincFlow/MS-GWaM with \cref{eq:leading-order-fluxes-msgwam,eq:tracer-advection-flux-Q0}, we conduct two test cases examining the impact of a three-dimensional gravity wave packet on the large-scale tracer distribution. In the first case (\cref{sec:q0-rest}), a gravity wave propagates through an initially resting atmosphere and has an initial aspect ratio of \(\lambda_y/\lambda_z=\num{300}\), where \(\lambda_y\) and \(\lambda_z\) are the horizontal and vertical wavelengths, respectively. The second case (\cref{sec:q0-jet}) introduces a background jet and uses a gravity wave with an initial aspect ratio \(\lambda_y/\lambda_z=\num{10}\). These contrasting setups illustrate that the parameterization performs consistently across different background flows and wave configurations.

In both test cases, a high-resolution wave-resolving simulation serves as our reference and will be referred to as 3DWR. 
To ensure the gravity wave is well resolved, we use at least 16 grid points per initial wavelength in each horizontal direction and 10 grid points per wavelength in the vertical direction.
Additionally, we perform a coarse-resolution where the resolution is approximately one grid point per wavelength. Since the gravity wave impact, including \(\mathcal{Q}^{(0)}\), is not resolved at this resolution, it is parameterized using MS-GWaM. We denote this simulation as 3DCR. 
Finally, to assess the effect of the tracer flux convergence in \cref{eq:tracer-advection-flux-Q0}, we conduct another simulation with parameterized waves, but with \(\mathcal{Q}^{(0)}\) neglected. This simulation is referred to as 3DCRno. As \(\mathcal{Q}^{(0)}\) is the largest gravity wave-effect on tracer transport in these cases, both 3DCR and 3DCRno do not include the next-order effect \(\mathcal{Q}^{(1)}\).

\subsubsection{Wave packet in an atmosphere at rest} \label{sec:q0-rest}

The first case we present is of a gravity wave packet in a rotating, isothermal atmosphere at rest, with a temperature of \(T_0=300\text{K}\). A locally monochromatic wave packet is introduced, where the atmospheric variables are given by 
\begin{align}
    \vb{v} &= \vb{v}' \;, \label{eq:initial_wind} \\
    \theta &= \overline{\theta} + \theta' \;, \\
    \psi &= \mean{\psi} + \psi' \;.
\end{align}
The potential temperature perturbation is initialized via the buoyancy using the WKB ansatz 
\begin{equation} \label{eq:theta_prime}
    \theta' = \frac{\overline{\theta}}{g} b'(\vb{x},t) = \frac{\overline{\theta}}{g}\mathfrak{R}\left[\hat{b}(\vb{x},t)e^{i(l_0y+m_0z)}\right] \;,
\end{equation}
with wave amplitude \(\hat{b}\) and initial wavenumbers \(l_0\) and \(m_0\) in the \(y\)- and \(z\)-directions, respectively. 
The initial perturbations in the wind \(\vb{v}'\) and tracer mixing ration \(\psi'\) are determined from \(b'\) using the polarization relations (\cref{eq:polarization-u,eq:polarization-w,eq:amplitude-eq-full-gamma0}). The buoyancy wave amplitude is initially defined by a Gaussian function in the \(z\)-direction and a cosine function in the \(x\)-direction, with no dependence on \(y\):
\begin{equation}
    \label{eq:3d-wavepacket-buoyancy-amplitude}
    \hat{b}(\vb{x},t=0) = a_0\frac{N^2}{m_0^2}\exp\left[-\frac{(z-z_0)^2}{2\sigma_z^2} \right]
    \begin{cases}
        \cos\left[\frac{\pi(x-x_0)}{2\sigma_x} \right] & \text{for} \; |x-x_0|\leq \sigma_x \;, \\
        0 & \text{else}\;,
    \end{cases}
\end{equation}
where \((x_0,z_0)\) is the position of the wave amplitude maximum in the \(x\)-\(z\)-plane, while \(\sigma_x\) and \(\sigma_z\) define the width of the wave packet in the \(x\)- and \(z\)-directions, respectively. The initial vertical wavenumber \(m_0\) is chosen such that \(m_0<0\), ensuring the wave packet propagates upwards when selecting the positive branch of the intrinsic frequency and setting \(k_0>0\). The factor \(a_0\) scales the buoyancy amplitude relative to the threshold of static instability, which is reached at \(a_0=1\). Additionally, the initial large-scale tracer distribution \(\mean{\psi}\) is prescribed as a linear function of altitude:
\begin{equation} \label{eq:initial-large-scale-tracer}
    \mean{\psi} 
    (t=0)
    = 
    \alpha_{\psi}z \;,
\end{equation}
where \(\alpha_{\psi}\) is an arbitrary scaling parameter. All simulation parameters are summarized in \cref{tab:simulation-parameters}.

\begin{table}[ht]
\centering
\caption{Initial model parameters for the simulations of the three-dimensional wave packet in an initially resting atmosphere (3D), in an atmosphere with a jet (3Djet), and the two-dimensional wave packet (2D). The number of grid points \(N_x\), \(N_y\), and \(N_z\) are given for the wave-resolving simulations (WR) and the coarse-resolution simulations with parameterized gravity waves (CR).}\label{tab:simulation-parameters}
\begin{threeparttable}
\begin{tabular}{c c c c}
\headrow
\thead{Parameter} & \thead{3D} & \thead{3Djet} & \thead{2D} \\
Wavelength in \(x\)-direction \(\lambda_x\) [km] & -- & -- & \(1\)  \\
Wavelength in \(y\)-direction \(\lambda_y\) [km] & \(300\) & \(30\) & --  \\
Wavelength in \(z\)-direction \(\lambda_z\) [km] & \(-1\) & \(-3\) & \(-1\)  \\
Branch of intrinsic frequency \(\hat{\omega}\) & \(+1\) & \(+1\) & \(+1\)  \\
Wave amplitude factor \(a_0\) & \(0.5\) & \(0.5\) & \(0.1\)  \\
Width of wave packet in \(x\)-direction \(\sigma_x\) [km] & \(1500\) & \(150\) & -- \\
Width of wave packet in \(z\)-direction \(\sigma_z\) [km] & \(5\) & \(5\) & \(2\) \\ 
Wave packet position in \(x\)-direction \(x_0\) [km] & \(4500\) & \(400\) & -- \\ 
Wave packet position in \(z\)-direction \(z_0\) [km] & \(30\) & \(10\) & \(10\) \\
Jet strength \(u_0\) [ms\(^{-1}\)] & --  & \(10\) & -- \\
Width of jet in \(z\)-direction \(z_0\) [km] & --  & \(10\) & -- \\
Jet position in \(z\)-direction \(z_0\) [km] & --  & \(10\) & -- \\
Coriolis parameter \(f\) [s\(^{-1}\)] & \(10^{-4}\) & \(10^{-4}\) & \(0\) \\
Brunt-Vaisala frequency \(N\) [s\(^{-1}\)] & \(0.02\) & \(0.02\) & \(0.02\) \\
Atmosphere temperature \(T_0\) [K] & 300 & 300 & 300 \\
Tracer mixing ratio factor \(\alpha_{\psi}\) [m\(^{-1}\)] & 1 & 1 & 1 \\
Domain size in \(x\)-direction \(L_x\) [km] & \(9000\) & \(800\) & \(1\) \\
Domain size in \(y\)-direction \(L_y\) [km] & \(300\) & \(30\) & -- \\
Domain size in \(z\)-direction \(L_z\) [km] & \(100\) & \(100\) & \(30\) \\
Number of grid points in \(x\)-direction (WR) \(N_x\) & \(512\) & \(460\) & 32 \\
Number of grid points in \(y\)-direction (WR) \(N_y\) & \(16\) & \(16\) & -- \\
Number of grid points in \(z\)-direction (WR) \(N_z\) & \(1000\) & \(1000\) & 960 \\
Number of grid points in \(x\)-direction (CR) \(N_x\) & \(32\) & \(32\) & 3 \\
Number of grid points in \(y\)-direction (CR) \(N_y\) & \(1\) & \(1\) & -- \\
Number of grid points in \(z\)-direction (CR) \(N_z\) & \(100\) & \(100\) & 300 \\
\hline
\end{tabular}
\end{threeparttable}
\end{table}

In the coarse-resolution simulations 3DCR and 3DCRno, the large-scale tracer distribution \(\mean{\psi}\) and velocity \(\mean{\vb{v}}\)are obtained directly as output. However, in the high-resolution wave-resolving simulation 3DWR, the output consists of the full variables \(\psi\) and \(\vb{v}\), which include both the large-scale and gravity wave components. Therefore, post-processing is required to extract \(\mean{\psi}\) and \(\mean{\vb{v}}\). Since the initial wave packet setup in \cref{eq:3d-wavepacket-buoyancy-amplitude} assumes no dependence on the \(y\)-coordinate for the large-scale variables, we can extract them by averaging the full variables over one wavelength in the \(y\)-direction. The gravity wave components are then determined as 
\begin{align}
    \psi' &= \psi - \mean{\psi} \;, \label{eq:psi_prime}\\
    \vb{v}' &= \vb{v} - \mean{\vb{v}} \;. \label{eq:v_prime}
\end{align}
This enables us to compute the leading-order gravity wave tracer flux convergence 
\begin{equation}
 \mathcal{Q}^{(0)} = - \frac{1}{\overline{\rho}}\nabla\cdot\mean{\overline{\rho}\vb{v}'\psi'} \;,
\end{equation}
which we can then compare to the \(\mathcal{Q}^{(0)}\) obtained from MS-GWaM.

The results after 1500 min are visualized in \cref{fig:2DWP90_comparison}, showing the gravity wave tracer flux convergence \(\mathcal{Q}^{(0)}\) and the resulting change in the tracer distribution for 3DWR, 3DCR, and 3DCRno. 
In the reference simulation 3DWR, the gravity wave packet leads to a decrease in the tracer mixing ratio to the left and an increase to the right of the wave packet amplitude maximum. 
This behaviour can be explained by the initial setup of \(\vb{k}_0=(0,l_0,m_0)\) and \(\nabla\mean{\psi} = \alpha_{\psi}\vb{e}_z\), as the wave packet has changed only slightly during the 1500-minute simulation due to its relatively small vertical group velocity. 
Comparing 3DCR to the reference simulation 3DWR, we find that the fluxes and the resulting changes in the large-scale tracer distribution are well reproduced. To assess the effect of setting \(\mathcal{Q}^{(0)}=0\), we compare 3DCR to 3DCRno, where the latter neglects \(\mathcal{Q}^{(0)}\). In 3DCRno, the large-scale tracer \(\mean{\psi}\) is only advected by the wave-induced large-scale wind (not shown). The resulting \(\Delta\mean{\psi}\) differs significantly from that in 3DCR, and consequently from 3DWR.
Thus, we conclude that the changes in \(\Delta\mean{\psi}\) observed in 3DWR and 3DCR are significantly impacted by \(\mathcal{Q}^{(0)}\) and not merely a result of advection due to the large-scale wind \(\mean{\vb{v}}\).

\subsubsection{Wave packet in an atmosphere with a jet} \label{sec:q0-jet}

The second test case for validating of the computation of \(\mathcal{Q}^{(0)}\) initializes a rotating, isothermal (\(T_0=\SI{300}{\kelvin}\)) atmosphere with a jet defined by
\begin{equation}
    \mean{\vb{v}}(\vb{x},t=0) = 
    \begin{cases}
    \frac{u_{0}}{2}\left\{1 + \cos\left[\frac{\pi \left(z-z_{\text{jet}}\right)}{\sigma_{\text{jet}}}\right]\right\}\vb{e}_x  &\text{for} \; |z-z_{\text{jet}}| \leq \sigma_{\text{jet}}\\
    0 & \text{else} \;.
    \end{cases}
\end{equation}
Because of the Coriolis effect, the jet's axis rotates counter-clockwise in the \(x\)-\(y\)-plane over an \SI{18}{\hour} period. The large-sacle tracer distribution is initialized following \cref{eq:initial-large-scale-tracer}. A gravity-wave packet, specified by \cref{eq:theta_prime,eq:3d-wavepacket-buoyancy-amplitude}, is introduced with its maximum located at the vertical center of the jet. The simulation parameters are summarized in \cref{tab:simulation-parameters}. 

\begin{figure}[t]
 \noindent\includegraphics[width=\textwidth,angle=0]{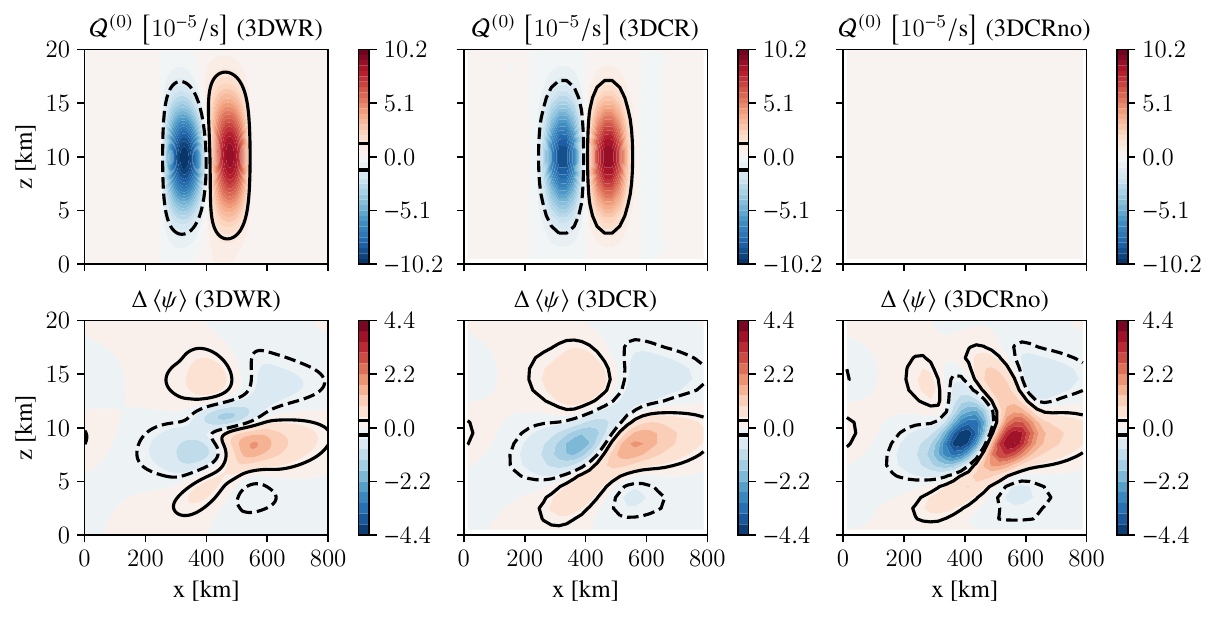}
 \caption{Gravity wave-induced tracer flux convergence \(\mathcal{Q}^{(0)}\) (top row) and change in the large-scale tracer distribution \(\Delta\mean{\psi}\) (bottom row) from a three-dimensional gravity wave packet with a jet centered at \(z=\SI{10}{\kilo\meter}\), at \(t=\SI{6}{\hour}\); (left column) 3DWR: wave-resolving reference simulation; (center column) 3DCR: coarse-resolution simulation with parameterized gravity wave and including \(\mathcal{Q}^{(0)}\); (right column) 3DCRno: coarse-resolution simulation with \(\mathcal{Q}^{(0)}=\mathcal{Q}^{(1)}=0\). The solid and dashed contour lines illustrate the regions with positive and negative values, respectively.}\label{fig:3DWP_jet}
\end{figure}

\Cref{fig:3DWP_jet} shows the tracer flux convergence \(\mathcal{Q}^{(0)}\) and the resulting six-hour change in the large-scale tracer field, \(\Delta\mean{\psi}\), for the three simulations, 3DWR, 3DCR, and 3DCRno. Computing \(\mathcal{Q}^{(0)}\) with MS-GWaM enables the coarse-resolution run (3DCR) to reporduce the \(\Delta\mean{\psi}\) seen in the wave-resolving reference (3DWR), with only minor differences attributable to resolution. In contrast, when the Stokes drift contribution is omitted (3DCRno), the coarse-resolution simulation exhibits strong deviations from the reference. Even in the presence of background shear flows, including the leading-order gravity-wave impact is essential for accurate tracer-transport simulations. 

Taken together with the first test case (\cref{sec:q0-rest,sec:q0-jet}), these results demonstrate that the computation of \(\mathcal{Q}^{(0)}\) performs reliably across different background flows and initial wavelength settings. 

\subsection{Parameterization of the Next-order Gravity Wave Tracer Flux Convergence} \label{sec:next-order-results}

\begin{figure}[t]
 \noindent\includegraphics[width=\textwidth,angle=0]{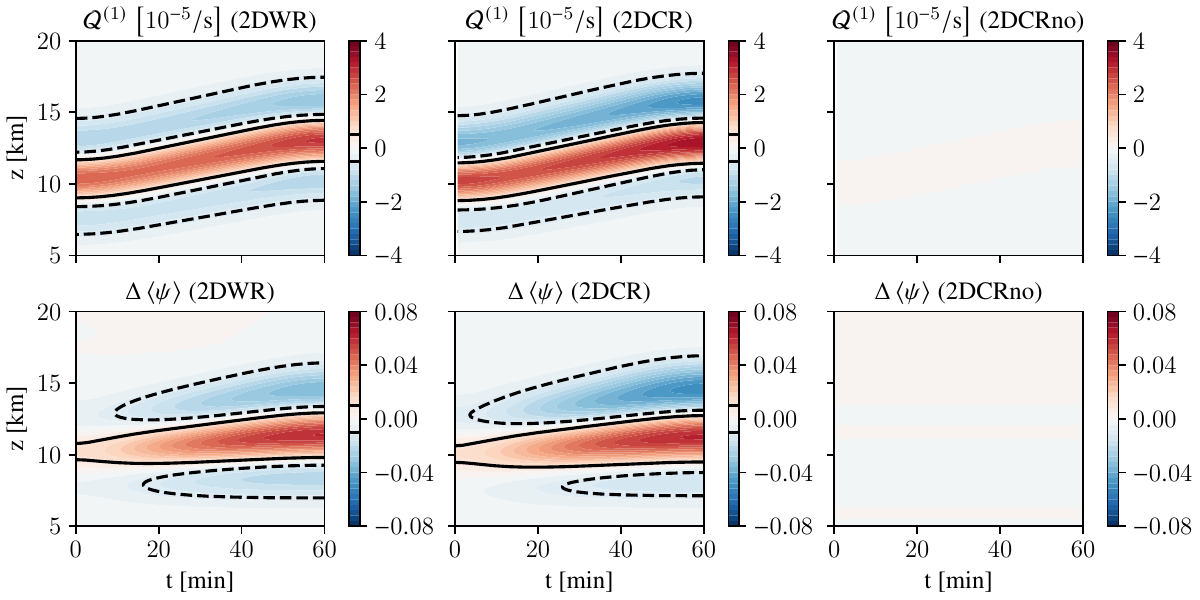}
 \caption{Time evolution of the gravity wave-induced tracer flux convergence \(\mathcal{Q}^{(1)}\) (top row) and change in the large-scale tracer distribution \(\Delta\mean{\psi}\) (bottom row) from a two-dimensional gravity wave packet; (left column) 2DWR: wave-resolving reference simulation; (center column) 2DCR: coarse-resolution simulation with parameterized gravity wave and including \(\mathcal{Q}^{(1)}\); (right column) 2DCRno: coarse-resolution simulation with \(\mathcal{Q}^{(0)}=\mathcal{Q}^{(1)}=0\). The solid and dashed contour lines illustrate the regions with positive and negative values, respectively.}\label{fig:STINH_comparison}
\end{figure}

In the following, we present the extension of MS-GWaM to calculate the next-order flux convergence \(\mathcal{Q}^{(1)}\). To validate this extension, we conduct simulations of a vertically propagating, two-dimensional, locally monochromatic gravity wave packet in an isothermal, non-rotating atmosphere. The absence of rotation in the atmosphere ensures that there are no leading-order effect induced by inertial gravity waves.

From the extension of MS-GWaM to calculate \(\mathcal{Q}^{(1)}\), we assume a locally monochromatic wave packet. Dropping the index \(\alpha\) from the equations summarized in \cref{sec:summary-equations}, the wave energy is given 
\begin{align}
 E_w = \overline{\rho}\frac{\left|\hat{b}\right|^2}{2N^2}\frac{\hat{\omega}^2\left(k_h^2+m^2\right)}{N^2k_h^2} = \int \dd[3]{k}\hat{\omega}\mathcal{N} \;,
\end{align}
and we can therefore determine the absolute magnitude of the buoyancy
\begin{align}
 \left|\hat{b}\right| = \left[\int\dd[3]{k}\frac{\mathcal{N}}{\overline{\rho}}\frac{2N^4k_h^2}{\hat{\omega}\left(k_h^2+m^2\right)}\right]^{1/2} \;.
\end{align}
Within MS-GWaM, the wavenumber integral is calculated by summing the contributions from all ray volumes within a PincFlow finite volume cell. As discussed in \cref{sec:theory}, we can determine both the leading- and next-order wave amplitudes from the buoyancy amplitude \(\hat{b} = \left|\hat{b}\right|e^{i\varLambda}\), allowing us to parameterize the next-order tracer flux convergence. Numerically, all required temporal and spatial derivatives are approximated using simple finite differences. It is important to note that in the present monochromatic case, \cref{eq:ls_phase_dim} predicts that the large-scale phase remains invariant. 

The tracer transport equation is thus extended to 
\begin{equation} \label{eq:tracer-transport-q1}
    \pdv{(\rho\psi)}{t} + \nabla \cdot (\rho\psi\vb{v}) = \rho\mathcal{Q}^{(1)} \;,
\end{equation}
where \cref{eq:tracer-transport-q1} is used in coarse-resolution simulations with parameterized gravity waves through MS-GWaM, while wave-resolving simulations solve \cref{eq:tracer-advection-flux}.

We conduct a wave-resolving simulation (2DWR) as a reference simulation and a simulation with parameterized gravity waves (2DCR). To emphasize the impact of \(\mathcal{Q}^{(1)}\), we also conduct an additional coarse-resolution simulation with parameterized waves where we set \(\mathcal{Q}^{(1)}=0\) (2DCRno). In the case of 2DWR, the resolution is 16 grid points per wavelength in the \(x\)-direction and 10 grid points per vertical wavelength in the \(z\)-direction. For 2DCR and 2DCRno, we choose one grid point per wavelength in both the horizontal and vertical directions. Furthermore, as the atmosphere is nonrotating, \(\mathcal{Q}^{(0)}=0\) for both 2DCR and 2DCRno.

We initialize the isothermal atmosphere (\(T_0=300\text{K}\)) with a two-dimensional wave packet, where the buoyancy amplitude depends only on altitude and is given by a Gaussian function,
\begin{equation}
    \hat{b} = a_0\frac{N^2}{m_0^2}\exp\left[-\frac{(z-z_0)^2}{2\sigma^2}\right] \;.
\end{equation}
The buoyancy perturbation \(b'\) is therefore
\begin{equation}
    b' = \Re\left[\hat{b}e^{i(k_0x+m_0z)}\right] \;,
\end{equation}
from which all remaining perturbations in the atmospheric variables are calculated using the polarization relations. Since the wave packet is meant to propagate upwards, we select the positive branch of the frequency with \(k_0>0\) and \(m_0<0\) for the horizontal and vertical wavenumbers, respectively. 
As we are interested only in the second-order gravity wave effects, and not in turbulent mixing induced by breaking gravity waves, we set \(a_0=0.1\) so that the gravity wave packet remains far from static instability throughout the simulation.
The initial large-scale tracer distribution is linearly increasing with altitude, as given in \cref{eq:initial-large-scale-tracer}.

We calculate the large-scale variables from the 2DWR results by averaging over one wavelength in the horizontal direction. From there, we can calculate the perturbations in the atmospheric variables, as given in  \cref{eq:psi_prime,eq:v_prime}. Using these, we can then calculate the tracer flux convergence 
\begin{equation}
    \mathcal{Q}^{(1)} = \frac{1}{\overline{\rho}}\pdv{z}\mean{\overline{\rho}w'\psi'} \;.
\end{equation}
As the atmosphere is not rotating, we can be sure that there are no effects due to leading-order fluxes, as those are of an inertial nature.

The results of 2DWR, 2DCR, and 2DCRno over a simulation time of 60 min are visualized in \cref{fig:STINH_comparison}. 
Both the next-order tracer flux convergence \(\mathcal{Q}^{(1)}\) and the change in the large-scale tracer distribution calculated in 2DWR are replicated by the parameterized gravity wave simulation. We observe that the gravity wave causes a vertical tracer flux towards the center of the wave packet. In the case of 2DCRno, where \(\mathcal{Q}^{(1)}=0\), we observe no change in the large-scale tracer distribution, as there is no large-scale vertical wind (not shown). We can therefore conclude that the \(\Delta\mean{\psi}\) seen in the results for 2DWR and 2DCR are accurately parameterized in 2DCR.

\section{Summary and conclusions} \label{sec:summary}

While the large-scale zonal-averaged transport of tracers is characterized by global circulations and well resolved in weather and climate models, the effect of small-scale processes, such as gravity waves must be parameterized. Therefore, this study investigates the impact of small-scale gravity waves on large-scale tracer transport and presents how this impact can be calculated within a gravity wave parameterization model. We employ a multiple-scale analysis of the governing equations, assuming that the large-scale flow and wave amplitudes vary slowly in space and time compared to the rapid fluctuations of gravity waves. The analysis relied on the scale separation parameter \(\varepsilon\), utilizing the WKB ansatz to describe gravity waves and systematically expanding the equations in powers of \(\varepsilon\). Sorting the resulting equations by their orders in \(\varepsilon\) enabled the identification of the most significant impacts of gravity waves on tracer transport.

Our findings reveal the leading-order and next-order contributions of small-scale gravity waves to large-scale tracer transport. The leading-order impact, induced by inertial gravity waves, acts perpendicular to both the large-scale tracer gradient and the wave number. The next-order impact becomes particularly significant at lower latitudes, where the Coriolis parameter becomes small and therefore also the leading-order impact, due to its non-zero contribution in the absence of rotation. These impacts scale as \(\mathcal{O}\left(\varepsilon\right)\) and \(\mathcal{O}\left(\varepsilon^2\right)\), respectively, when compared to the large-scale wind advection (\(\mathcal{O}(1)\)). Additionally, we demonstrated that gravity wave effects can be calculated from the leading-order buoyancy amplitude and gravity wave energy. We extended a gravity wave parameterization model to calculate these effects. Validation of these results was performed using three idealized wave packet test cases, where comparisons with wave-resolving reference simulations confirmed the accuracy of the approach.

These findings enable highly accurate simulations of tracer transport that incorporate gravity wave effects without the computational cost of resolving the waves explicitly. This work represents an important step toward a comprehensive understanding of the small-scale non-chemical processes that perturb tracer distributions. To build on this foundation, future research should aim to include the effects of diffusive mixing due to turbulence, such as that caused by breaking gravity waves.

\FloatBarrier

\section*{acknowledgements}
I. K. and U. A. thank the German Research Foundation (DFG) for partial support through CRC 301 “TPChange” (Project No. 428312742 and Projects B06 “Impact of small-scale dynamics on UTLS transport and mixing,” B07 “Impact of cirrus clouds on tropopause structure,” and Z03 “Joint model development and modelling synthesis”). U. A. thanks the German Research Foundation (DFG) for partial support through the CRC 181 “Energy transfers in Atmosphere and Ocean” (Project No. 274762653 and Projects W01 “Gravity-wave parameterization for the atmosphere” and S02 “Improved Parameterizations and Numerics in Climate Models”).

\section*{data statement}
The code and data described in this study are available on request.

\section*{conflict of interest}
The authors declare no conflicts of interest.

\printendnotes

\begin{appendix}
    
\section{Generalized Derivation of Gravity Wave-Induced Tracer Fluxes} \label{sec:appendix-full-expansion}

In \cref{sec:theory,sec:summary-equations}, we derived expressions for the gravity wave-induced tracer fluxes (\cref{eq:leading-order-flux-convergence-redim,eq:next-order-gw-forcing_dim}), summarized in dimensional form, under the simplifying assumption of a locally monochromatic, large-amplitude gravity wave in a weakly stratified atmosphere (\(\delta=1\)). In this appendix, we extend the derivation to a more general setting. Specifically, we consider strongly stratified atmosphere (\(\delta=0\)), weakly nonlinear and quasi-linear gravity wave fields, and spectra of waves including higher harmonics.

For a spectrum of gravity waves with \(A\) wavenumbers, the atmospheric variables can be decomposed into three contributions: the reference atmosphere, the large-scale flow, and the wave field:
\begin{align}
    \label{eq:tracer-decomp-app}
    \psi(\vb{x},t) =   & \sum_{m=0}^{\infty}\varepsilon^m\mean{\Psi}^{(m)}(\cxv,\ct)
    + \mathfrak{R}\sum_{\alpha=1}^{A}\sum_{\beta=1}^{\infty}\sum_{m=0}^{\infty}\varepsilon^m\hat{\Psi}_{\alpha\beta}^{(m)}(\cxv,\ct)e^{i\beta\phi_{\alpha}(\cxv,\ct)/\varepsilon} \;,                                                                                                       \\
    \label{eq:wind-decomp-app}
    \vb{v}(\vb{x},t) = & \sum_{m=0}^{\infty} \varepsilon^m\mean{\vb{V}}(\cxv,\ct) + \varepsilon^{\gamma}\mathfrak{R}\sum_{\alpha=1}^{A}\sum_{\beta=1}^{\infty}\sum_{m=0}^{\infty}\varepsilon^{m}\hat{\vb{V}}_{\alpha\beta}^{(m)}(\cxv,\ct)e^{i\beta\phi_{\alpha}(\cxv,\ct)/\varepsilon} \;, \\
    \label{eq:pot-temp-decomp-app}
    \begin{split}
        \theta(\vb{x},t) =& \sum_{m=0}^{\delta}\varepsilon^m\overline{\Theta}^{(m)}(\cz) + \varepsilon^{\delta+1}\sum_{m=0}^{\infty}\varepsilon^{m}\mean{\Theta}^{(m)}(\cxv,\ct) \\
        &+ \varepsilon^{\gamma+\delta+1}\mathfrak{R}\sum_{\alpha=1}^{A}\sum_{\beta=1}^{\infty}\sum_{m=0}^{\infty}
        \varepsilon^{m}\hat{\Theta}_{\alpha\beta}^{(m)}(\cxv,\ct)e^{i\beta\phi_{\alpha}(\cxv,\ct)/\varepsilon} \;,
    \end{split}                                                                                                                \\
    \label{eq:exner-decomp-app}
    \begin{split}
        \pi(\vb{x},t) =& \sum_{m=0}^{\delta}\varepsilon^m\overline{\Pi}^{(m)}(\cz) +
        \varepsilon^{\delta+1}\sum_{m=0}^{\infty}\varepsilon^m\mean{\Pi}^{(m)}(\cxv,\ct) \\
        &+ \varepsilon^{\gamma+\delta+2}\mathfrak{R}\sum_{\alpha=1}^{A}\sum_{\beta=1}^{\infty}\sum_{m=0}^{\infty}\varepsilon^m\hat{\Pi}_{\alpha\beta}^{(m)}(\cxv,\ct)e^{i\beta\phi_{\alpha}(\cxv,\ct)/\varepsilon} \;.
    \end{split}
\end{align}
The wave terms are proportional to \(e^{i\beta\phi_{\alpha}/\varepsilon}\). Here, \(\alpha\) indexes distinct waves, each with its own phase function \(\phi_{\alpha}\). The harmonic index \(\beta\) refers to integer multiples of the same phase, so that \(\beta\)-th harmonic of wave \(\alpha\) has phase \(\beta\phi_{\alpha}\). In other words, different phase structures belong to different waves (different \(\alpha\)), while harmonics of a given wave share the same underlying phase function. The corresponding wavenumbers and frequencies are denoted by \(\beta\vb{k}_{\alpha}=\beta\nabla_{\cxv}\phi_{\alpha}\) and \(\beta\omega_{\alpha}=-\beta\partial_{\ct}\phi_{\alpha}\), respectively.

The scaling factor \(\varepsilon^{\gamma}\) distinguishes three regimes:
\begin{itemize}
    \item Large-amplitude waves (\(\gamma=0\)): Locally monochromatic case with \(A=1\).
    \item Weakly nonlinear waves (\(\gamma=1\)): Spectrum of waves allows (\(A\geq 1\)), but wave-wave interactions affecting the gravity waves themselves are neglected.
    \item Quasi-linear waves (\(\gamma=2\)): No restrictions.
\end{itemize}
We impose no amplitude assumptions on the tracer field, whose magnitude is determined directly from the governing equations. 

These assumptions are essential to keep the governing equations solvable. Higher-order amplitude terms introduce wave-wave interactions, which render the system nonlinear in the wave amplitudes and prohibitively complex. Since our goal is to construct a system of equations suitable for parameterization, we exclude such terms. This simplification is also physically justified, as the essential impact of gravity waves on tracer fluxes is preserved.

A summary of the indices \(\alpha\), \(\beta\), \(\gamma\), and \(\delta\) is given in \cref{tab:indices}.

The derivations of the tracer flux convergences \(\mathcal{Q}^{(0)}\) and \(\mathcal{Q}^{(1)}\) proceed analogously to \cref{sec:theory}. Here we present only the generalized form and highlight the main differences. The leading-order tracer convergence becomes:
\begin{equation}
    \label{eq:leading-order-gw-forcing-app}
    \mathcal{Q}^{(0)}\equiv -\frac{\varepsilon^{(2\gamma+2)}}{2\overline{P}}\sum_{\alpha=1}^{A}\nabla_{\cxv}\cdot\mathfrak{R}\left(\overline{P}\hat{\vb{V}}_{\alpha 1}^{(0)}\hat{\Psi}_{\alpha 1}^{(\gamma+1)^*}\right) \;.
\end{equation}
The next-order flux convergence is:
\begin{equation}
    \label{eq:next-order-gw-forcing-app}
    \begin{split}
        \mathcal{Q}^{(1)} \equiv -\frac{\varepsilon^{(2\gamma+3)}}{2\overline{P}}\sum_{\alpha=1}^{A}&\nabla_{\cxv}\cdot\mathfrak{R}\left[\overline{P}\left(\hat{\vb{V}}_{\alpha 1}^{(0)}\hat{\Psi}_{\alpha 1}^{(\gamma+2)^*} + \hat{\vb{V}}_{\alpha 1}^{(1)}\hat{\Psi}_{\alpha 1}^{(\gamma+1)^*}\right)\right]  \;.
    \end{split}
\end{equation}
Only the fundamental wave (\(\beta=1\)) contributes to the flux convergence, since \(\hat{\vb{V}}_{\alpha\beta}^{(0)}=0\) for all \(\beta>1\) \citep{Achatz2023MultiScaleDynamics}. Consequently, the tracer amplitude \(\hat{\Psi}_{\alpha\beta}^{(\gamma+1)}\) vanishes for higher harmonics as
\begin{equation}
    \label{eq:amplitude-eq-full-gamma0-app}
    i\beta\hat{\omega}_{\alpha}\hat{\Psi}_{\alpha\beta}^{(\gamma+1)} =
    \hat{\vb{V}}_{\alpha\beta}^{(0)}\cdot\nabla_{\cxv}\mean{\underline{\Psi}}\;.
\end{equation}
The intrinsic frequency is denoted by \(\hat{\omega}_{\alpha}=\omega_{\alpha}-\vb{k}_{h,\alpha}\cdot\mean{\underline{\vb{V}}}\). To obtain \(\mathcal{Q}^{(0)}\), we employ the wind amplitude relations, modified by replacing in \cref{eq:wind0,eq:wind1} the amplitude \(\hat{\vb{V}}^{(m)}\) with \(\hat{\vb{V}}_{\alpha 1}^{(m)}\) for \(m=0,1\), and \(\vb{k}\) with \(\vb{k}_{\alpha}\). For \(\mathcal{Q}^{(1)}\), we use the next-order tracer amplitude relation 
\begin{equation}
    \label{eq:amplitude-eq-full-gamma1-app}
    i\beta\hat{\omega}_{\alpha}\hat{\Psi}_{\alpha\beta}^{(\gamma+2)} = \left(\partial_{\ct} +
    \mean{\underline{\vb{U}}}\cdot\nabla_{\cxv,h}\right)
    \hat{\Psi}_{\alpha\beta}^{(\gamma+1)} + \hat{\vb{V}}_{\alpha\beta}^{(1)}\cdot\nabla_{\cxv}\mean{\underline{\Psi}}\;.
\end{equation}
As noted, wave-wave interactions in the weakly nonlinear regime (\(\gamma=1\)) would appear as triad terms in \cref{eq:amplitude-eq-full-gamma1-app}. These are neglected here to keep the problem tractable. Similarly, in the strongly stratified case (\(\delta=0\)), the correction terms 
\begin{equation}
    i\frac{c_V}{R}\frac{i \hat{\omega}_{\alpha}}{\underline{\overline{\Pi}}}\hat{\Pi}^{(0)}_{\alpha 1} - \frac{\mean{\underline{\Pi}}}{\overline{\underline{\Pi}}}\frac{1}{\mean{P}}\nabla\cdot\left(\mean{P}\hat{\vb{V}}_{\alpha 1}^{(0)}\right)+\frac{\mean{\underline{\Pi}}}{\overline{\underline{\Pi}}}\frac{1}{
        \overline{P}
    }
    \nabla_{\cxv}\cdot\left(
    \overline{P}
    \hat{\vb{V}}_{\alpha 1}^{(0)}\right) \;,
\end{equation}
appear on the right-hand side of the relation for the wind amplitude
\begin{equation} \label{eq:wind2}
    i\vb{k}_{\alpha}\cdot\hat{\vb{V}}_{\alpha 1}^{(2)}
    = -\frac{1}{
        \overline{P}
    }
    \nabla_{\cxv}\cdot\left(
    \overline{P}
    \hat{\vb{V}}_{\alpha 1}^{(1)}\right)  \;,
\end{equation}
with \(\mean{P}=\mean{\rho}\mean{\theta}\), which we neglect for tractability.
For a spectrum of waves, the leading-order flux convergence can be written as
\begin{equation}
    \mathcal{Q}^{(0)} = - \frac{\varepsilon^{(2\gamma+1)}}{\overline{\rho}}\sum_{\alpha=1}^{A}\nabla_{\cxv}\cdot
    \left(
    \frac{f_0}{\hat{\omega}_{\alpha}}\frac{m_{\alpha}}{\varepsilon^{5-\delta}k_{h,\alpha}^2+m_{\alpha}^2}\mathcal{A}_{\alpha}
    \vb{k}_{\alpha}\times\nabla_{\cxv}\mean{\underline{\Psi}}
    \right)
    \;,
\end{equation}
with wave action density \(\mathcal{A}_{\alpha}=E_{w,\alpha}/\hat{\omega}_{\alpha}\), wave energy \(E_{w,\alpha}\)
\begin{equation}
    \label{eq:wave-energy-app}
    E_{w,\alpha} = \overline{\rho} \frac{\left|\hat{B}_{\alpha 1}^{(0)}\right|^2}{2 N_0^2}
    \frac{N_0^2k_{h,\alpha}^2+f_0^2m_{\alpha}^2}{N_0^2 k_{h,\alpha}^2} \;,
\end{equation}
and the leading-order buoyancy amplitude \(\hat{B}_{\alpha 1}^{(0)}=\hat{\Theta}_{\alpha 1}^{(0)}/\overline{\underline{\Theta}}\). For the calculation of \(\mathcal{Q}^{(1)}\), we first decompose the buoyancy amplitude 
\begin{equation}
    \label{eq:buoyancy-decomp-app}
    \hat{B}_{\alpha 1}^{(0)} = \left|\hat{B}_{\alpha 1}^{(0)}\right|e^{i\varLambda_{\alpha}} \;.
\end{equation}
The absolute magnitude follows from the wave energy in \cref{eq:wave-energy-app}, while the large-scale phase \(\varLambda_{\alpha}\) evolves according to the prognostic equation:
\begin{equation}
    \label{eq:wave-phase-equation-app}
    \begin{split}
        \left(\partial_{\ct} + \vb{c}_{g,\alpha}\cdot\nabla_{\cxv}\right)\varLambda_{\alpha}
        =&
        - \frac{f_0m_{\alpha}}{2\left|\vb{k}_{\alpha}\right|^2\hat{\omega}_{\alpha}}
        \left(\vb{k}_{\alpha}\times\nabla_{\cxv}\right)\mean{\underline{\vb{U}}} \\
        &+ \frac{|\vb{k}_{\alpha}|^2}{2m_{\alpha}\left(N_0^2-\varepsilon^{5-\delta}f_0^2\right)}
        \left[\frac{1}{\overline{\underline{\Theta}}}\left(\hat{\vb{c}}_{g,\alpha}\cdot\nabla_{\cxv}\right)\mean{\underline{\Theta}}
        -\varepsilon^{\delta}N_0^2\frac{c_p}{R}\overline{\underline{\Theta}}\left(\hat{\vb{c}}_{g,\alpha}\cdot\nabla_{\cxv}\right)\mean{\underline{\Pi}}
            \right] \\
        &- \frac{f_0\hat{\omega}_{\alpha}m_{\alpha}^2}{k_{h,\alpha}^2\left(N_0^2-\varepsilon^{5-\delta}f_0^2\right)}
        \left[\left(l_{\alpha}\partial_{\cx}-k_{\alpha}\partial_{\cy}\right)\hat{\omega}_{\alpha}+l_{\alpha}\left(\partial_{\ct}+\mean{\underline{\vb{U}}}\cdot\nabla_{\cxv,h}\right)k_{\alpha}
        - k_{\alpha}\left(\partial_{\ct}+\mean{\underline{\vb{U}}}\cdot\nabla_{\cxv,h}\right)l_{\alpha}\right] \;,
    \end{split}
\end{equation}
with group velocity \(\vb{c}_{g,\alpha}=\nabla_{\vb{k}}\omega_{\alpha}\) and intrinsic group velocity \(\hat{\vb{c}}_{g,\alpha}=\nabla_{\vb{k}}\hat{\omega}_{\alpha}\). Next, we solve the matrix equation:
\begin{equation}
    \label{eq:matrix-eq-1-app}
    M_{\alpha 1}\vb{Z}_{\alpha 1}^{(1)} = \vb{R}_{\alpha 1}^{(1)} \;,
\end{equation}
with matrix
\begin{equation}
    \label{eq:matrix-app}
    M_{\alpha 1}=M(\vb{k}_{\alpha},\hat{\omega}_{\alpha}) =
    \mqty[-i\hat{\omega}_{\alpha} & -f_0 & 0 & 0 & i k_{\alpha} \\
        f_0 & -i\hat{\omega}_{\alpha} & 0 & 0 & i l_{\alpha} \\
        0 & 0 & -i\hat{\omega}_{\alpha}\varepsilon^{5-\delta} & -N_0 & i m_{\alpha} \\
        0 & 0 & N_0 & -i\hat{\omega}_{\alpha} & 0 \\
        i k_{\alpha} & i l_{\alpha} & i m_{\alpha} & 0 & 0] \;,
\end{equation}
next-order wave amplitude vector
\begin{equation}
    \vb{Z}_{\alpha 1}^{(1)} = \left(
    \hat{U}_{\alpha 1}^{(1)}\:,\:
    \hat{V}_{\alpha 1}^{(1)}\:,\:
    \hat{W}_{\alpha 1}^{(1)}\:,\:
    \hat{B}_{\alpha 1}^{(1)}/N_0\:,\:
    \frac{c_p}{R}
    \overline{\underline{\Theta}}
    \hat{\Pi}_{\alpha 1}^{(1)}\right)^{\mathrm{T}} \;,
\end{equation}
and right-hand side vector
\begin{equation}
    \vb{R}_{\alpha 1}^{(1)} = \left(
    R_{u,\alpha 1}\:,\:
    R_{v,\alpha 1}\:,\:
    R_{w,\alpha 1}\:,\:
    R_{b,\alpha 1}/N_0\:,\:
    R_{\pi,\alpha 1}\right)^{\mathrm{T}} \;.
\end{equation}
The components of \(\vb{R}_{\alpha 1}^{(1)}\) are
\begin{align}
    \begin{split}
        \vb{R}_{\vb{u},\alpha 1}^{(1)} =& - \left(\partial_{\ct} + \mean{\underline{\vb{U}}}\cdot\nabla_{\cxv,h}\right)\hat{\vb{U}}_{\alpha 1}^{(0)}
        - \left(\hat{\vb{V}}_{\alpha 1}^{(0)}\cdot\nabla_{\cxv}\right)\mean{\underline{\vb{U}}}
        - \frac{c_p}{R}\left[
            \overline{\underline{\Theta}} \nabla_{\cxv,h}\hat{\Pi}_{\alpha 1}^{(0)} + \varepsilon^\delta
            \left(i\vb{k}_{h,\alpha}\mean{\underline{\Theta}}
            \hat{\Pi}_{\alpha 1}^{(0)}+ \hat{\Theta}_{\alpha 1}^{(0)}\nabla_{\cxv,h}\mean{\underline{\Pi}} \right)\right] \;,
    \end{split} \\
    \begin{split}
        R_{w,\alpha 1}^{(1)} =& -\varepsilon^{5-\delta}\left(\partial_{\ct} + \mean{\underline{\vb{U}}}\cdot\nabla_{\cxv,h}\right)\hat{W}_{\alpha 1}^{(0)}
        - \frac{c_p}{R}\left[\overline{\underline{\Theta}} \partial_{\cz}\hat{\Pi}_{\alpha 1}^{(0)} +
            \varepsilon^\delta
            \left(im_{\alpha}\mean{\underline{\Theta}}\hat{\Pi}_{\alpha 1}^{(0)} + \hat{\Theta}_{\alpha 1}^{(0)}
            \partial_{\cz}\mean{\underline{\Pi}}\right)\right] \;,
    \end{split}            \\
    \begin{split}
        R_{b,\alpha 1}^{(1)} =& -\left(\partial_{\ct} + \mean{\underline{\vb{U}}}\cdot\nabla_{\cxv,h}\right)
        \hat{B}_{\alpha 1}^{(0)}
        - \frac{\hat{\vb{V}}_{\alpha 1}^{(0)}}{\overline{\underline{\Theta}}}
        \cdot\nabla_{\cxv}\mean{\underline{\Theta}} \;,
    \end{split}                                                           \\
    \begin{split}
        R_{\pi,\alpha 1}^{(1)}
        =&
        - \frac{1}{\overline{P}}
        \nabla_{\cxv}\cdot
        \left(
        \overline{P}
        \hat{\vb{V}}_{\alpha 1}^{(0)}
        \right) \;,
    \end{split}
\end{align}
where \(\vb{R}_{\vb{u},\alpha 1}^{(1)} = \left(R_{u,\alpha 1}^{(1)}\:,\:R_{v,\alpha 1}^{(1)}\right)^{\mathrm{T}}\). The leading-order wave amplitudes are obtained from \(\hat{B}_{\alpha 1}^{(0)}\) using the polarization relations
\begin{align}
    \label{eq:polarization-u-app}
    \hat{\vb{U}}_{\alpha 1}^{(0)}                                        & = \frac{i}{m_{\alpha}N_0^2}\frac{\varepsilon^{5-\delta}\hat{\omega}_{\alpha}^2-N_0^2}{\hat{\omega}_{\alpha}^2-f_0^2}\left(\vb{k}_{h,\alpha}\hat{\omega}_{\alpha}-if_0\vb{e}_z\times\vb{k}_{h,\alpha}\right)\hat{B}_{\alpha 1}^{(0)}\;, \\
    \label{eq:polarization-w-app}
    \hat{W}_{\alpha 1}^{(0)}                                             & = \frac{i\hat{\omega}_{\alpha}}{N_0^2}\hat{B}_{\alpha 1}^{(0)} \;,                                                                                                                                                                     \\
    \label{eq:polarization-pi-app}
    \frac{c_p}{R}\underline{\overline{\Theta}}\hat{\Pi}_{\alpha 1}^{(0)} & =
    \frac{i}{m_{\alpha}}\frac{\varepsilon^{5-\delta}\hat{\omega}_{\alpha}^2-N_0^2}{N_0^2}
    \hat{B}_{\alpha 1}^{(0)} \;.
\end{align}
Finally, with \cref{eq:amplitude-eq-full-gamma1-app}, the next-order flux convergence in \cref{eq:next-order-gw-forcing-app} can be evaluated.

In dimensional form, the leading-order tracer flux convergence is:
\begin{equation} \label{eq:leading-order-flux-convergence-redim-app}
    \mathcal{Q}^{(0)} = -\frac{1}{\overline{\rho}}\nabla\cdot\sum_{\alpha=1}^{A}\frac{f}{\hat{\omega}_{\alpha}}\frac{m_{\alpha}}{k_{h,\alpha}^2+m_{\alpha}^2}\mathcal{A}_{\alpha}
    \vb{k}_{\alpha}\times\nabla\mean{\psi}
    \;,
\end{equation}
while the next-order flux convergence is:
\begin{equation}
    \label{eq:next-order-gw-forcing_dim-app}
    \begin{split}
        \mathcal{Q}^{(1)}
        =
        -\frac{1}{2 \overline{\rho}\overline{\theta}}
        \sum_{\alpha=1}^{A}&\nabla\cdot\mathfrak{R}
        \left[
            \overline{\rho}\overline{\theta}
            \left(
            \vb{v}_\alpha'\psi_{\alpha}''^* + \vb{v}_\alpha''\psi_{\alpha}'^*
            \right)\right] \;.
    \end{split}
\end{equation}
The calculation of \(\mathcal{Q}^{(1)}\) proceeds sequentially:
\begin{enumerate}
    \item Solve for the leading-order buoyancy amplitude \(b_{\alpha 1}'=\left|b_{\alpha 1}'\right|e^{i\varLambda_{\alpha}}\) from the wave energy equation
          \begin{equation}
              E_{w,\alpha}
              = \frac{\overline{\rho}}{2}\left(\frac{\left|\vb{v}_{\alpha}'\right|^2}{2} + \frac{\left|b_{\alpha}'\right|^2}{2N^2} \right)
              = \overline{\rho} \frac{\left|b_\alpha'\right|^2}{2 N^2} \frac{N^2k_{h,\alpha}^2+f^2m_{\alpha}^2}{N^2 k_{h,\alpha}^2} \;.
          \end{equation}
          and integrate the prognostic equation for \(\varLambda_{\alpha}\) to obtain the large-scale phase:
          \begin{equation}
              \label{eq:ls_phase_dim-app}
              \begin{split}
                  \left(\partial_t + \vb{c}_{g,\alpha}\cdot\nabla\right)\varLambda_{\alpha}
                  =&
                  - \frac{fm_{\alpha}}{2\left|\vb{k}_{\alpha}\right|^2\hat{\omega}_{\alpha}}\left(\vb{k}_{\alpha}\times\nabla\right)\mean{\vb{u}} \\
                  &+ \frac{|\vb{k}_{\alpha}|^2}{2m_{\alpha}\left(N^2-f^2\right)}
                  \left[\frac{g}{\overline{\theta}}\left(\hat{\vb{c}}_{g,\alpha}\cdot\nabla\right)\mean{\theta} -\frac{N^2}{g}c_p\overline{\theta}\left(\hat{\vb{c}}_{g,\alpha}\cdot\nabla\right)\mean{\pi}
                      \right] \\
                  &- \frac{f\hat{\omega}_{\alpha}m_{\alpha}^2}{k_{h,\alpha}^2\left(N^2k_{h,\alpha}^2+f^2m_{\alpha}^2\right)}
                  \left[\left(l_{\alpha}\partial_x-k_{\alpha}\partial_y\right)\hat{\omega}_{\alpha}
                  + l_{\alpha}\left(\partial_t+\mean{\vb{u}}\cdot\nabla_h\right)k_{\alpha}
                  - k_{\alpha}\left(\partial_t+\mean{\vb{u}}\cdot\nabla_h\right)l_{\alpha}\right] \;.
              \end{split}
          \end{equation}
    \item Compute the leading-order wind and Exner pressure amplitudes using the polarization relations:
          \begin{align}
              \label{eq:polarization-u-dim-app}
              \vb{u}_{\alpha}'                  & = \frac{i}{m_{\alpha}N^2}\frac{\hat{\omega}_{\alpha}^2-N^2}{\hat{\omega}_{\alpha}^2-f^2}\left(\vb{k}_{h,\alpha}\hat{\omega}_{\alpha}-if\vb{e}_z\times\vb{k}_{h,\alpha}\right)b_{\alpha}' \;, \\
              \label{eq:polarization-w-dim-app}
              w_{\alpha}'                       & = \frac{i\hat{\omega}_{\alpha}}{N^2}b_{\alpha}' \;,                                                                                                                                          \\
              \label{eq:polarization-pi-dim-app}
              c_p\overline{\theta}\pi_{\alpha}' & = \frac{i}{m_{\alpha}}\frac{\hat{\omega}_{\alpha}^2-N^2}{N^2}b_{\alpha}' \;.
          \end{align}
    \item Evaluate the right-hand side terms:
          \begin{align}
              \label{eq:rhs_u_dim-app}
              \begin{split}
                  \vb{R}_{\vb{u},\alpha} =& -\left(\partial_t + \mean{\vb{u}}\cdot\nabla_h
                  \right)\vb{u}_{\alpha}'
                  - \vb{v}_\alpha'
                  \cdot\nabla\mean{\vb{u}}
                  - c_p
                  \left[
                      \overline{\theta}\nabla_h
                      \pi_\alpha'
                      + i\vb{k}_{h,\alpha}\mean{\theta}\pi_\alpha'+\theta_{\alpha}'\nabla_h\mean{\pi}
                      \right]\;,
              \end{split}              \\
              \label{eq:rhs_w_dim-app}
              \begin{split}
                  R_{w,\alpha} =& -\left(\partial_t + \mean{\vb{u}}\cdot\nabla_h
                  \right)w_{\alpha}' - c_p
                  \left[
                      \overline{\theta}\partial_z\pi_{\alpha}'
                      + im_{\alpha}\mean{\theta}\pi_{\alpha}' + \theta_\alpha'\partial_z\mean{\pi}
                      \right]\;,
              \end{split} \\
              \label{eq:rhs_b_dim-app}
              \begin{split}
                  R_{b,\alpha}
                  =&
                  -\left(\partial_t + \mean{\vb{u}}\cdot\nabla_h
                  \right)
                  b_\alpha'
                  - \frac{\vb{v}_\alpha'}{\overline{\theta}}
                  \cdot\nabla\mean{\theta} \;,
              \end{split}                                                    \\
              \label{eq:rhs_pi_dim-app}
              \begin{split}
                  R_{\pi,\alpha}
                  =& -\frac{1}{
                      \overline{\rho}
                      \overline{\theta}}
                  \nabla\cdot
                  \left(
                  \overline{\rho}
                  \overline{\theta}
                  \vb{v}_\alpha'
                  \right) \;.
              \end{split}
          \end{align}
          and solve the matrix equation for \(\vb{Z}''_{\alpha}\):
          \begin{equation}
              \label{eq:matrix-eq-1-redim-app}
              \underbrace{\mqty[-i\hat{\omega}_{\alpha} & -f & 0 & 0 & ik_{\alpha} \\
                      f & -i\hat{\omega}_{\alpha} & 0 & 0 & il_{\alpha} \\
                      0 & 0 & -i\hat{\omega}_{\alpha} & -N & im_{\alpha} \\
                      0 & 0 & N & -i\hat{\omega}_{\alpha} & 0 \\
                      ik_{\alpha} & il_{\alpha} & im_{\alpha} & 0 & 0]}_{M_{\alpha}}
              \underbrace{\mqty[u_{\alpha}'' \\ v_{\alpha}'' \\ w_{\alpha}'' \\
                      b_{\alpha}''/N \\ c_p\overline{\theta}\pi_{\alpha}'']}_{\vb{Z}_{\alpha}''} =
              \underbrace{\mqty[R_{u,\alpha} \\ R_{v,\alpha} \\ R_{w,\alpha} \\
                      R_{b,\alpha}/N \\ R_{\pi, \alpha}]}_{\vb{R}_{\alpha}} \;,
          \end{equation}
    \item Determine the leading- and next-oder tracer amplitudes from:
          \begin{align}
              \label{eq:tracer_lead_dim-app}
              \psi_{\alpha}' & = -\frac{i}{\hat{\omega}_{\alpha}}\vb{v}_{\alpha}'\cdot\nabla\mean{\psi} \;.
          \end{align}
          and
          \begin{equation}
              \label{eq:tracer_next_dim-app}
              i\hat{\omega}_{\alpha}
              \psi_{\alpha}''
              =
              -\left(\partial_t + \mean{\vb{u}}\cdot\nabla_h
              \right)
              \psi_{\alpha}'
              + \vb{v}_{\alpha}''
              \cdot\nabla\mean{\psi} \;.
          \end{equation}
    \item Substitute these amplitudes and the corresponding wind fields into \cref{eq:next-order-gw-forcing_dim-app} to obtain \(\mathcal{Q}^{(1)}\).
\end{enumerate}

This generalized derivation extends the results of \cref{sec:theory} and \cref{sec:summary-equations} to cover spectra of gravity waves, large-amplitude, weakly nonlinear, and quasi-linear regimes, and both weakly and strongly stratified atmospheres. While simplified for clarity, the formulation provides a flexible basis for parameterizing tracer fluxes under realistic atmospheric conditions.

\begin{table}[]
    \label{tab:indices}
    \centering
    \caption{Summary of the indices that represent various wave cases and atmospheric conditions.}
    \begin{threeparttable}
        \begin{tabular}{l l l}
            \headrow
            \thead{Index} & \thead{Description}                                         & \thead{Values }                                                                                                    \\
            \(\alpha\)    & Wave modes: indicates the wave in a spectrum of waves with \(A\) wavenumbers & \(\{1...A\}\)                                                                                                      \\
            \(\beta\)     & Higher harmonics of the wave                                & \begin{tabular}[c]{@{}l@{}}\(1\): basic wave\\ \(>1\): higher harmonic\end{tabular}                                \\
            \(\gamma\)    & Measure of non-linearity of the gravity waves               & \begin{tabular}[c]{@{}l@{}}\(0\): high-amplitude wave\\ \(1\): weakly nonlinear\\ \(2\): quasi-linear\end{tabular} \\
            \(\delta\)    & Stratification of the reference atmosphere                  & \begin{tabular}[c]{@{}l@{}}\(0\): strong\\ \(1\): weak\end{tabular}
        \end{tabular}
    \end{threeparttable}
\end{table}
\end{appendix}

\bibliography{sample}


\end{document}